\newif\ifTWO
\global\TWOtrue

\newif\ifONE
\ifTWO
\documentclass[preprint2]{aastex}
\global\ONEfalse
\else
\documentclass[manuscript]{aastex}
\global\ONEtrue
\fi

\usepackage{epsfig}
\usepackage{url}
\usepackage{lineno}

\usepackage{color}    
\newcommand{\tcx}{\textcolor{black}}

\newcommand{\info}{Appendix}

\ifTWO
\newcommand{\bls}{1.0}
\newcommand{\blsA}{1.0}
\newcommand{\blsC}{1.0}
\else
\newcommand{\bls}{2.0}
\newcommand{\blsA}{1.5}
\newcommand{\blsC}{1.0}
\fi

\newcommand{\D}{\Delta}
\newcommand{\Dt}{\mbox{$\D t$}}
\newcommand{\eM}{\mbox{$e_{\cal M}$}}
\newcommand{\DeM}{\mbox{$\D e_{\cal M}$}}
\newcommand{\aM}{\mbox{$a_{\cal M}$}}
\newcommand{\mM}{\mbox{$m_{\cal M}$}}
\newcommand{\aV}{\mbox{$a_{\cal V}$}}
\newcommand{\mV}{\mbox{$m_{\cal V}$}}
\newcommand{\ms}{{\tt mercury6}}
\newcommand{\hnb}{{\tt HNBody}}
\newcommand{\e}[1]{\mbox{$\x10^{#1}$}}
\newcommand{\x}{\times}
\newcommand{\bm}{\boldmath}
\newcommand{\ubm}{\unboldmath}
\renewcommand{\v}[1]{\mbox{\bm$#1$\ubm}}
\newcommand{\beqn}{\begin{eqnarray}}
\newcommand{\eeqn}{\end{eqnarray}}
\newcommand{\beq}{\begin{eqnarray*}}
\newcommand{\eeq}{\end{eqnarray*}}
\newcommand{\pmo}{\mbox{$^{-1}$}}
\newcommand{\q}{\frac}
\newcommand{\nn}{\nonumber}
\newcommand{\eE}{\mbox{$e_{\cal E}$}}
\newcommand{\aE}{\mbox{$a_{\cal E}$}}
\newcommand{\iE}{\mbox{$i_{\cal E}$}}
\newcommand{\dm}{\mbox{$d_{min}$}}
\newcommand{\rH}{\mbox{$r_H$}}
\newcommand{\BS}{Bulirsch-Stoer}
\newcommand{\tBS}{\mbox{$t_{\rm BS}$}}
\newcommand{\MV}{$\cal M$-$\cal V$}
\newcommand{\MS}{$\cal M$-$\cal S$}
\newcommand{\at}{\mbox{$\widetilde{a}$}}
\newcommand{\Et}{\mbox{$\widetilde{E}$}}
\newcommand{\Lt}{\mbox{$\widetilde{\v{L}}$}}

\def\NSY{\#0649}
\def\NBS{\#0299}
\def\RSY{R0649}
\def\RBS{R0299}
\def\RXZ{R0043} 
\def\RDE{R0728} 
\def\RXA{R1461} 

\def\figdir{}

\shorttitle{Earth's future orbit}
\shortauthors{Zeebe}

\begin{document}

\ifONE
\linenumbers
\fi

\title{Highly stable evolution of Earth's future orbit 
despite chaotic behavior of the Solar System}

\author{Richard E. Zeebe$^{1,*}$}
\affil{\vspace*{0.5cm}
     $^*$Corresponding Author.\\ 
     $^1$School of Ocean and Earth Science and Technology, 
     University of Hawaii at Manoa, 
     1000 Pope Road, MSB 629, Honolulu, HI 96822, USA. 
     email: zeebe@soest.hawaii.edu\\[2ex]
     Accepted, {\bf The Astrophysical Journal}\\ 
     \today \\
     }

\renewcommand{\baselinestretch}{\blsA}\selectfont
\begin{abstract}
Due to the chaotic nature of the Solar System, the question 
of its dynamic long-term stability can only be answered in a 
statistical sense, for instance, based on numerical ensemble 
integrations of nearby orbits. 
Destabilization of the inner planets, including catastrophic 
encounters and/or collisions involving the Earth, has been suggested
to be initiated through a large increase in 
Mercury's eccentricity (\eM), with an estimated probability 
of $\sim$1\%.
However, it has recently been shown that the statistics of numerical
Solar System integrations are sensitive to the accuracy and type 
of numerical algorithm. 
Here I report results from computationally demanding ensemble 
integrations ($N = 1,600$ with slightly different initial conditions) 
at unprecedented accuracy based on the full equations of 
motion of the eight planets and Pluto over 5~Gyr, including 
contributions from general relativity. 
The standard symplectic algorithm used for long-term integrations 
produced spurious results for highly eccentric orbits
and during close encounters, which were hence integrated with a 
suitable \BS\ algorithm, specifically designed for these situations.
The present study yields odds for a large increase in 
Mercury's eccentricity that are less than previous
estimates.
Strikingly, in two solutions Mercury continued on highly 
eccentric orbits (after reaching \eM\ values $>$0.93) for 
80-100~Myr before colliding with Venus or the Sun.
Most importantly, none of the 1,600 solutions led to
a close encounter involving the Earth or
a destabilization of Earth's orbit in the future.
I conclude that Earth's orbit is dynamically highly stable
for billions of years in the 
future, despite the chaotic behavior of the Solar System.
\end{abstract}


\renewcommand{\baselinestretch}{\bls}\selectfont
\section{Introduction}

One of the oldest and still active research areas in celestial 
mechanics is the question of the long-term dynamic stability 
of the Solar System. After centuries of analytical work led by
Newton, Lagrange, Laplace, Poincar{\'e}, Kolmogorov, Arnold,
Moser etc. \citep{laskar13}, the field has recently 
experienced a renaissance due to advances in numerical algorithms
and computer speed. Only since the 1990s have
researchers been able to integrate the full equations of motion
of the Solar System over time scales approaching its lifetime 
($\pm\sim$5~Gyr)
\citep{wisdom91,quinn91,sussman92,saha92,murray99,
ito02,varadi03,batygin08,laskar09,zeebe15apj},
and exceeding Earth's future habitability of perhaps 
another 1-3~Gyr \citep{schroeder08,rushby13}.
Looking ahead, detailed exoplanet observations 
\citep[e.g.][]{oppenheimer13} will enable similar 
integrations of planetary systems beyond our own 
solar system.

Importantly, in addition to CPU speed, a statistical approach is 
necessary due to the chaotic behavior of the Solar System, i.e.\ 
the sensitivity of orbital solutions to initial conditions
\citep{laskar89,sussman92,murray99,richter01,varadi03,batygin08,
laskar09,zeebe15apj}.
To obtain adequate statistics, ensemble integrations
are required \citep{laskar09,zeebe15apj}, that is, simultaneous 
integration of a large number of nearby orbits, e.g.\
utilizing massive parallel computing. Chaos in the
Solar System means
small differences in initial conditions grow exponentially, 
with a time constant (Lyapunov time, e.g.\
\citet{morbidelli02})
for the inner planets of $\sim$3-5~Myr estimated
numerically \citep{laskar89,varadi03,batygin08,zeebe15apj}
and $\sim$1.4~Myr analytically \citep{batygin15}. 
For example, a difference in initial position
of $1$~cm grows to $\sim$1~AU (=~1.496\e{11}~m) after 
90-150~Myr, which makes it fundamentally impossible to 
predict the evolution of planetary orbits accurately 
beyond $\sim$100~Myr \citep{laskar89}. 
Thus, rather than searching for a single deterministic 
solution (conclusively describing the Solar System's 
future evolution, see Laplace's demon, \citet{laplace51}), 
the stability question must be answered in probabilistic 
terms, e.g.\ by studying the behavior of a large number of 
physically possible solutions.

\tcx{
Until present, only a single study is
available that integrated a large number of Solar System
solutions based on the full equations of motion and
including contributions from general relativity
\citep{laskar09}. Using a symplectic integrator throughout, 
\citet{laskar09} integrated 2,501 orbits over 5~Gyr and 
found that Mercury's orbit achieved large eccentricities 
($>$0.9) in about 1\% of the solutions. Furthermore,
they suggested the possibility of a collision between 
the Earth and Venus. Given that only one study of this 
kind exists to date, several questions remain. 
For instance, are the numerical results and statistics
obtained sensitive to the numerical algorithm used in the 
integrations? Furthermore, is the symplectic algorithm used 
throughout a reliable integrator for highly eccentric orbits and 
during close encounters? If not, what are the consequences 
for Solar System stability? In particular, considering long-term
stability and planetary habitability, what are the 
consequences for the evolution of Earth's future orbit
over billions of years? 
Addressing these questions is the focus of the present study.
} 

\section{Methods}

I have integrated 1,600 solutions with slightly
different initial conditions \tcx{for Mercury's 
position} based on the full 
equations of motion of the 
eight planets and Pluto over 5~Gyr into the future using 
the numerical integrator packages \hnb\ \citep{rauch02} 
and \ms\ \citep{chambers99}. Relativistic corrections 
\citep{einstein16} are critical \citep{laskar09,zeebe15apj}
and were available in \hnb\ but not in \ms. Post-Newtonian 
corrections \citep{soffel89,poisson14} were therefore 
implemented before using \ms\ (see \info). Thus, all simulations 
presented here include contributions from general relativity 
(GR). To allow comparison with previous studies, higher-order 
effects such as asteroids \citep{ito02,batygin08},
perturbations from passing stars, and solar mass loss
\citep{ito02,batygin08,laskar09} were also not included 
here. 

The initial 5-Gyr integrations of the 1,600-member ensemble
were performed with \hnb\ \citep{rauch02} using a 4th-order 
symplectic integrator plus corrector with constant 4-day 
timestep (\Dt, Table~\ref{TabDtEcc}). The ensemble computations
required $\sim$6 weeks uninterrupted
wall-clock time on a Cray CS300 ($\sim$1.7 million core 
hours total), plus up to $\sim$4 months for individual 
runs at reduced timestep (see below). 
For the symplectic integrations, Jacobi coordinates 
\citep{wisdom91} were employed rather than heliocentric
coordinates, \tcx{as the latter}
may underestimate the odds for destabilization of 
Mercury's orbit \citep{zeebe15apj}. 

All simulations 
started from the same set of initial conditions 
(Table~\ref{TabDE431}), 
except Mercury's initial radial distance was offset by 
1.75~mm between every two adjacent orbits. The largest 
overall offset was $\rm1,599\x1.75~mm\simeq2.80$~m,
well within the uncertainty of our current knowledge of 
the Solar System. 
Initial conditions for all bodies
in the 5-Gyr runs (before offsetting Mercury)
were generated from DE431
(\url{naif.jpl.nasa.gov/pub/naif/generic_kernels/spk/planets})
at JD2451544.5 (01 Jan 2000) using the SPICE toolkit for Matlab 
(\url{naif.jpl.nasa.gov/naif/toolkit.html}) 
(Table~\ref{TabDE431}).
The small offsets in Mercury's initial position randomized 
the initial conditions and led to complete divergence of 
trajectories after $\sim$100~Myr 
\tcx{(see Fig.~\ref{FigEmax}).}

In case Mercury's eccentricity increased above certain threshold 
values during the simulation (Table~\ref{TabDtEcc}), \Dt\ was 
reduced but held constant after that until the next threshold
was crossed, if applicable.
\tcx{For example, a 4-fold reduction in stepsize, twice 
throughout the simulation (i.e.\ $4 \rightarrow 1 \rightarrow 
1/4$~day) typically kept 
the maximum relative error in energy $|\D E/E| = |(E(t)-E_0)/E_0|$ 
and angular momentum ($|\D L/L|$) of the symplectic integrator 
below $10^{-10}$ and $10^{-11}$ (Fig.~\ref{FigSplice}).
These steps in \Dt\ corresponded to \eM\ thresholds
of about 0.55 and 0.70 (Table~\ref{TabDtEcc}). All results 
of the symplectic \hnb\ integrations were inspected and 
(if applicable) restarted manually with smaller \Dt\ at the
appropriate integration time using saved coordinates from 
the run with larger \Dt\ (automatic stepsize 
reduction was not possible because \hnb's source 
code is not available).
}

\ifTWO \def\tx{0ex} \else \def\tx{-2ex} \fi

\begin{table}[t]
\caption{Summary of the 5-Gyr simulations. \label{TabDtEcc}}
\vspace*{5mm}
\hspace*{-5mm}
\begin{tabular}{llcrcc}
\tableline\tableline
\eM \ $^a$        & Algorithm$^b$ & $\Dt$  & $N$      
                  & \multicolumn{2}{c}{\#Collisions$^c$} \\
                  &               & (days) &
                                  & \MV    & \MS \\
\hline     
$<1.00$           & 4th-sympl.    & 4      & 1600   &    &   \\
$>0.55$           & 4th-sympl.    & 1      &   28   &    &   \\
$\gtrsim0.70^d$   & 4th-sympl.    & 1/4    &   10   &    &   \\
$\gtrsim0.80^e$   & BS          & adaptive &   10   & 7  & 3 \\  
\tableline
\end{tabular}

\noindent {\small
$^a$\eM\ = Mercury's eccentricity.             \\[\tx]
$^b$4th-sympl. = 4-th order symplectic (\hnb). 
BS = \BS\ (\ms).                               \\[\tx]
$^c$\MV: Mercury-Venus, \MS: Mercury-Sun.      \\[\tx]
$^d$Switched to $1/4$~day--symplectic in case of 
large $|\D E/E|$ variations at $1$~day.        \\[\tx]
$^e$Switched to BS in case of \aM\ shifts
and/or large $|\D E/E|$ variations at 
$1/4$~day--symplectic.
}
\end{table}

Importantly, high \eM\ solutions \tcx{associated with
\aM\ shifts and/or large $|\D E/E|$ variations (usually 
during close encounters)} were integrated using the \BS\ 
(BS) algorithm with adaptive stepsize control of \ms, including
GR contributions (see \info) and specifically designed for 
these tasks \citep{chambers99} (the symplectic algorithm 
produced spurious results in these situations, \tcx{see 
Figs.~\ref{FigFail}, \ref{FigFailAll}).}
\tcx{Specifically, the symplectic \hnb\ integrations (1/4~day)
were inspected for \aM\ shifts and/or large $|\D E/E|$ 
variations and restarted with \ms's BS algorithm at the appropriate 
integration time using saved coordinates from the symplectic 
\hnb\ run.}
The only comparable 
study to date \citep{laskar09} integrated 2,501 orbits using a 
symplectic algorithm and an initial $\Dt \simeq 
9$~days. While \citet{laskar09} also reduced the 
timestep depending on \eM\ (though with 
generally larger maximum errors, see below), high \eM\ 
solutions and close encounters were integrated using
the symplectic algorithm throughout \citep{laskar09}. 
However, for highly eccentric orbits the sympletic 
method can become unstable and may introduce artificial 
chaos, unless \Dt\ is small enough to always resolve 
periapse \citep{rauch99}. In the Solar System, \Dt\
must hence resolve Mercury's periapse with the
highest perihelion velocity ($v_p$) among the planets
(and increasing with $e$, as $v_p^2\propto (1+e)/(1-e)$).

\section{Results}

The vast majority of solutions obtained here showed a 
stable evolution of the Solar System and moderate
\eM\ values (\eM$<$0.6 in 99.3\% of all runs). In 10 out of 1,600 
solutions ($\sim$0.6\%), Mercury's eccentricity increased 
beyond 0.8 (see below). Importantly, the maximum relative error
in energy $|\D E/E|$ and angular
momentum ($|\D L/L|$) of the symplectic 
integrator was typically held below $10^{-10}$ 
and $10^{-11}$ even at $\eM \simeq 0.8$
by e.g.\ a 4-fold reduction in stepsize twice throughout
the simulation (Fig.~\ref{FigSplice}). Critically,
once \eM\ increased beyond $\sim$0.8 (resulting in close 
encounters with Venus and shifts in Mercury's semi-major 
axis, \aM), the \BS\ integrator of \ms\ was employed. 
For example, in run \NBS\ (\RBS\ for short) close encounters 
between Mercury and Venus 
\renewcommand{\baselinestretch}{\blsC}
\ifTWO
\begin{figure}[t]
\def\epsfsize#1#2{0.42#1}
\vspace*{-0mm}
\else
\begin{figure}[t]
\def\epsfsize#1#2{0.6#1}
\vspace*{-5mm}
\fi
\begin{center}
\centerline{\vbox{\epsfbox{\figdir 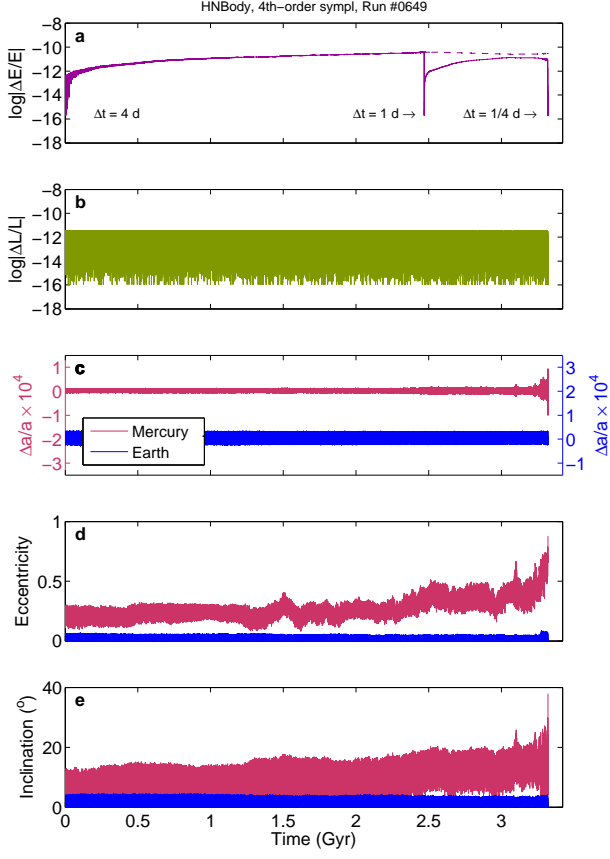}}}
\end{center}
\caption[]{\small Example of a symplectic multi-billion year
integration of the Solar System with \hnb\ (run \NSY).
Results shown are from symplectic integrator only
(4-th order with corrector and Jacobi coordinates),
not from \BS\ integrator (see text).
(a) Relative energy error, $|\D E/E| = |(E(t)-E_0)/E_0|$.
After the two reductions in integrator timestep ($t_i$,
$i = 1,2$), solid and dashed lines show $|\D E/E|$ relative to 
$E(t_i)$ and relative to absolute $E_0(t_0 = 0)$, respectively.
(b) Relative angular momentum error, $|\D L/L| = |(L(t)-L_0)/L_0|$. 
(c) Change in Mercury's and Earth's semi-major axes, 
$\D a/a = (a(t)-a_0)/a_0$.
(d) Mercury's and Earth's eccentricity (\eM, \eE). 
(e) Inclination. Note that $|\D E/E|
< 10^{-10}$, even at $\eM \simeq 0.8$. When oscillations 
in $|\D E/E|$ and/or shifts in \aM\ occurred
($\eM\gtrsim0.8$), 
integrations were continued with \ms's \BS\ algorithm 
\tcx{(see Figs.~\ref{FigFail}, \ref{FigBSm6}, 
\ref{FigFailAll}).}
}
\label{FigSplice}
\end{figure}
\ifONE \clearpage \fi
occurred at $t \simeq 4.3042$~Gyr
\tcx{at which time they approach} their mutual Hill radius 
\citep{chambers96}:
\beqn
r_{H} = \q{\aM + \aV}{2} \ 
       \left( \q{\mM + \mV}{3m_1} \right)^{1/3} 
      \simeq \rm 0.0053~AU \ ,
\eeqn
where $a$'s and $m$'s are the planetary semi-major axes 
and masses; $m_1$ is the solar mass.

\ifTWO
\begin{figure}[t]
\def\epsfsize#1#2{0.42#1}
\vspace*{-0mm}
\else
\begin{figure}[t]
\def\epsfsize#1#2{0.5#1}
\vspace*{-5mm}
\fi
\begin{center}
\centerline{\vbox{\epsfbox{\figdir 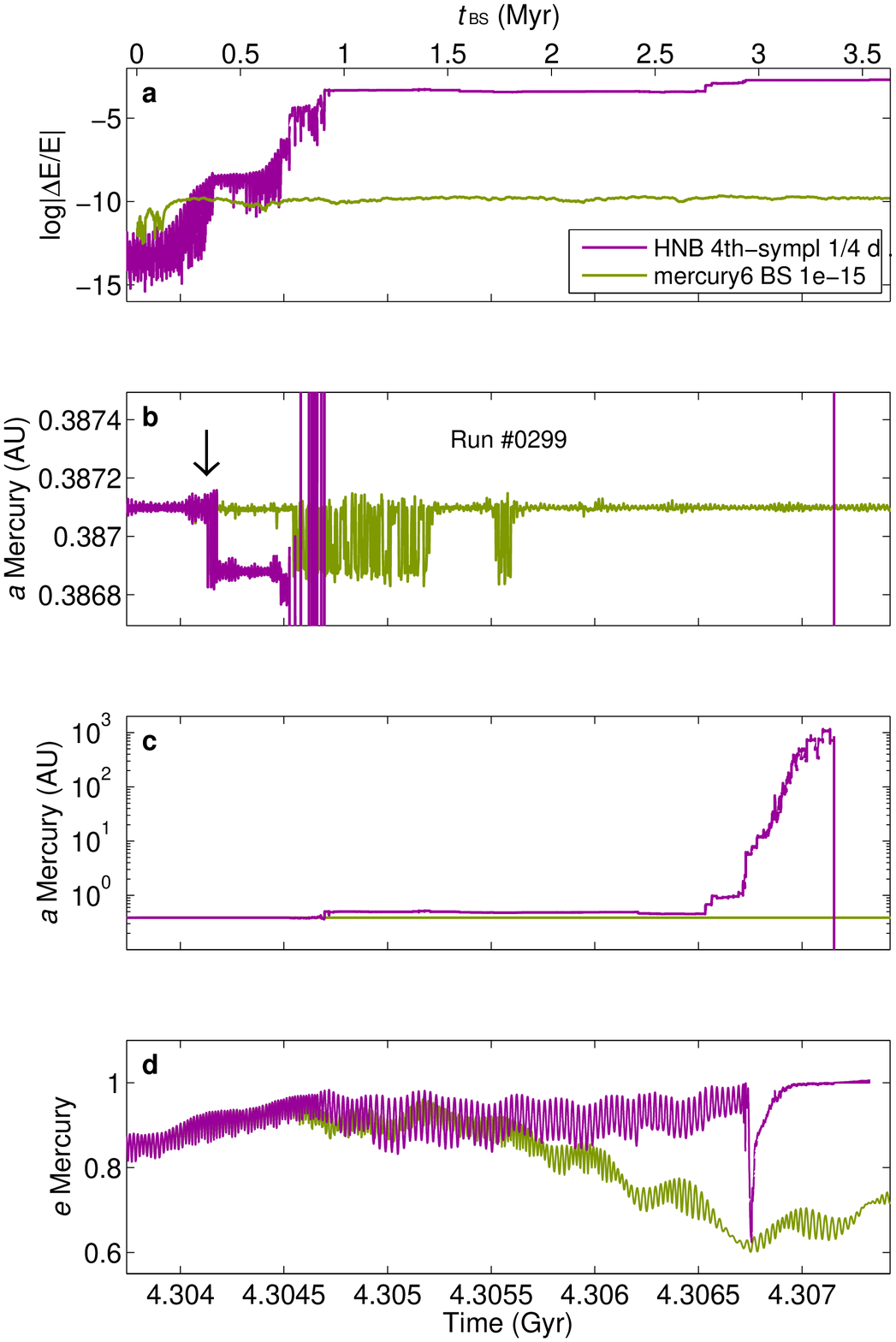}}}
\end{center}
\caption[]{\small Symplectic and \BS\ (BS) integration 
of run \NBS\ at high \eM. The BS integration (\ms, green lines,
approximate relative $\v{x}$ and $\v{v}$ error per step = $10^{-15}$)
was initiated at $\tBS = 0$ using the coordinates of the 
symplectic run (HNB = \hnb, purple lines, 4-th order, timestep 
= 1/4~day).
(a) Relative energy error, $|\D E/E|$. (b) and (c) Mercury's 
semi-major axis, \aM\ (note different y-scales). (d) Mercury's 
eccentricity, \eM.
Note spurious drop in \aM\ at $t_{BS}\simeq340$~kyr in 
the symplectic integration (arrow), which ultimately causes 
rapid destabilization of Mercury's orbit (see text).
In contrast, \aM\ is essentially stable in the BS integration, 
which properly resolves Mercury's periapse and close encounters
as a result of adaptive step size control.
}
\label{FigFail}
\end{figure}

\noindent
More precisely,
close encounters with $\dm = 0.0080$~AU ensued at 
$t_{BS}\simeq340$~kyr,
where $t_{BS}$ is measured from the start of the BS
integration and \dm\ is the minimum distance 
between the two bodies (Fig.~\ref{FigFail}). 
This point in time corresponds to a significant drop of
\aM\ to $\sim$0.3869~AU in the symplectic integration,
while \aM\ in the BS integration remains stable.
Note that the corresponding symplectic $|\D E/E| 
< 10^{-8}$ at 340~kyr may be misinterpreted to reflect 
accurate orbit integration \tcx{(in fact, if the symplectic 
\Dt\ was reduced right afterwards, spurious behavior would
likely go unnoticed)}. However, this is clearly not the
case as shown by comparison with BS, which,
due to adaptive step size control, properly resolves
Mercury's periapse and close encounters. Hence the
relative energy error (say $\lesssim10^{-8}$, 
\citet{laskar09}) 
is not a sufficient criterion to ensure accurate
steps in symplectic integrations with
highly eccentric orbits and close encounters.

\ifTWO
\begin{figure}[t]
\def\epsfsize#1#2{0.6#1}
\vspace*{0mm}
\hspace*{3cm}
\else
\begin{figure}[t]
\def\epsfsize#1#2{0.6#1}
\vspace*{-5mm}
\begin{center}
\fi
\centerline{\vbox{\epsfbox{\figdir 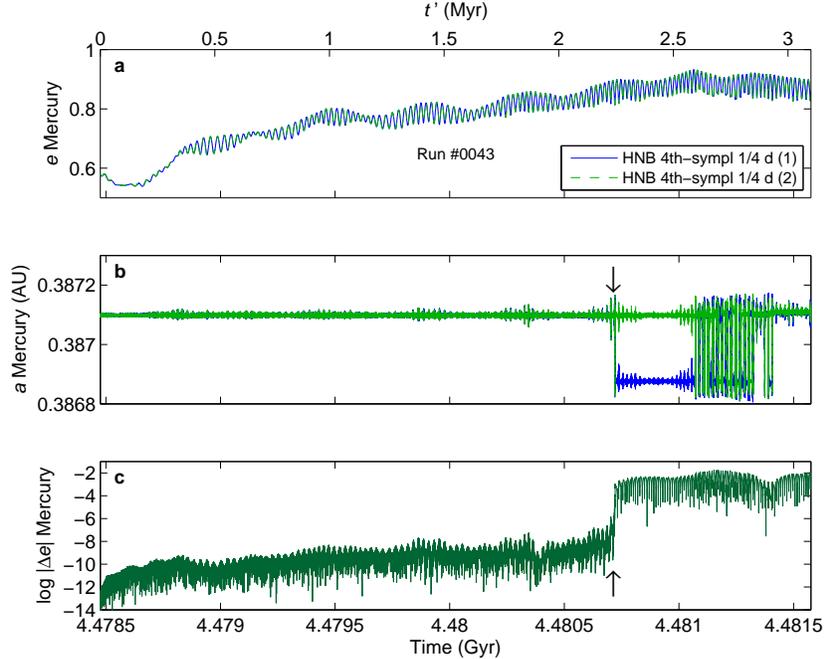}}}
\ifONE \end{center} \fi
\caption[]{\small
\tcx{
Illustration of trajectory separation as Mercury's 
orbit approaches instability. (a) \eM\ from two 
symplectic integrations with \hnb\ of Run 0043
of one fiducial (solid blue line, Nr.\ (1)) and one 
shadow orbit (dashed green line, Nr.\ (2)) offset 
by 1.75~mm in Mercury's $x$-coordinate at $t' = 0$.
(b) Mercury's semi-major axis. (c) Difference in
\eM\ (\DeM) between (1) and (2). At $t' \simeq 2.2$~Myr,
close encounters between Mercury and Venus ensue 
(see jumps in \aM\ and \DeM, arrows).
} 
}
\label{FigLyap}
\end{figure}

\ifTWO \newpage \vspace*{8.7cm} \noindent \fi

The drop in \aM\ at $t_{BS}\simeq340$~kyr in the symplectic
\RBS\ integration preconditions the system for further \eM\
increase, subsequently causing a large \aM\ rise and 
rapid destabilization of Mercury's 
orbit (Fig.~\ref{FigFail}). 
\tcx{Specifically, the \aM\ drop is followed by 
an immediate increase in the 
average \eM\ in the symplectic integration relative to BS,
which, in turn degrades Mercury's perihelion resolution 
(at constant symplectic \Dt) and leads to oscillations 
in $|\D E/E|$ and 
to further, larger swings in \aM. In contrast, while 
in the BS integration \aM\ also oscillates, it remains 
overall fairly stable over several million years.} 
Eventually, Mercury collides with Venus in BS-\RBS\
at $t_{BS}\simeq15$~Myr (Fig.~\ref{FigBSm6}). Thus, while ultimately 
both the symplectic
and BS integration of \RBS\ spell disaster for Mercury's 
orbit, it involves vastly different trajectories and time 
scales. Notably, Mercury's extended lifetime in the
BS integration relative to the symplectic algorithm
is a common feature \tcx{(see Figs.~\ref{FigBSm6},
\ref{FigFailAll}). One reason for this is of 
course a too large (and constant) timestep in
the current symplectic integrations. The crux, however, 
is that even if the symplectic timestep was reduced 
before $|\D E/E|$ becomes too large, spurious results 
could easily be overlooked.
}

Note that the spurious \aM\ shift in the symplectic 
integration of \RBS\ occurs after only
a few 100~kyr \tcx{during close encounters}
($t_{BS} = 0$ means same coordinates
in symplectic and BS integration) 
and is therefore not due to intrinsic chaos, 
\tcx{whose characteristic time scale causes trajectory
separation over millions of years. The million-year 
separation time still holds even as Mercury's orbit 
approaches the regime of instability in the solutions 
studied here ($\eM \gtrsim 0.8$). This behavior can be
illustrated by following two nearby trajectories 
initiated at high \eM\ (Fig.~\ref{FigLyap}).
One fiducial and one shadow orbit of R0043 offset 
by 1.75~mm in Mercury's $x$-coordinate
were integrated using \hnb's symplectic algorithm.
Polynomial growth in \DeM\ (difference in \eM\
between the two orbits) governs \DeM's increase 
over the first $\sim$2.2~Myr \citep[cf.][]{zeebe15apj}. 
However, at 
$t' \simeq 2.2$~Myr, close encounters between Mercury
and Venus ensue (as in \RBS), causing jumps in \aM\ and 
\DeM\ (arrows, Fig.~\ref{FigLyap}). Additional shadow
runs initiated at $t' = -26$~Myr and $-11$~Myr
(not shown) exhibit no close encounters during 
run-up to $t' = 2.2$~Myr and indicate exponential
trajectory divergence due to chaos after $\gtrsim 8$~Myr. 
Thus, the abrupt divergence of trajectories at 2.2~Myr 
is due to close encounters, not intrinsic chaos.
}

\ifTWO
\begin{figure}[t]
\def\epsfsize#1#2{0.40#1}
\hspace*{1.5cm}
\centerline{\vbox{\epsfbox{\figdir 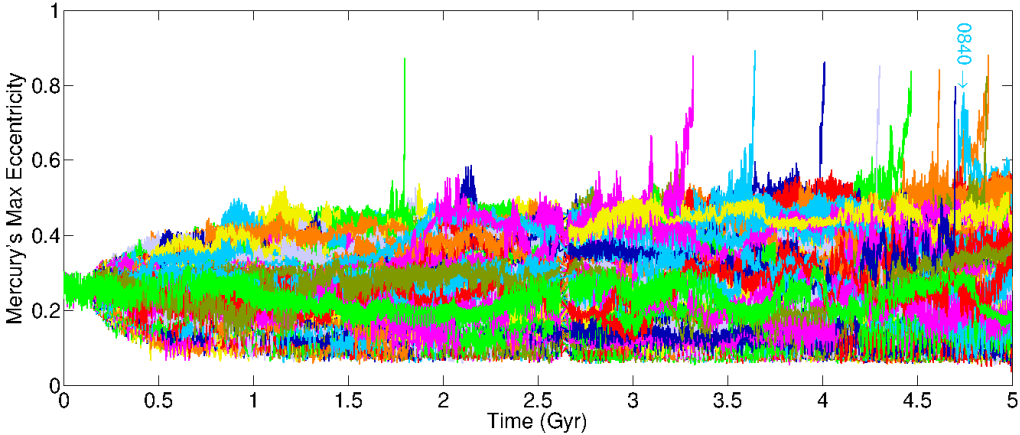}}}
\else
\begin{figure}[t]
\def\epsfsize#1#2{0.40#1}
\begin{center}
\centerline{\vbox{\epsfbox{\figdir emax.M.2.eps}}}
\end{center}
\fi
\caption[]{\small 
Mercury's maximum eccentricity per 1~Myr bin from the
1,600 symplectic 5-Gyr integrations. Results shown are from 
\hnb's symplectic integrator only \citep{rauch02}. 
In ten solutions, \eM\ crossed 0.8, after which the integration 
was continued with \ms's \BS\ algorithm \citep{chambers99} 
(Fig.~\ref{FigBSm6}). Note that in R0840 (arrow), $\eM < 0.8$, 
no shifts in \aM\ occurred, and $\max|\D E/E| \simeq 8\e{-11}$.
}
\label{FigEmax}
\end{figure}

\ifTWO \newpage \vspace*{6cm} \noindent \fi

\subsection{High-\eM\ solutions}

In ten solutions, \eM\ increased beyond 0.8 of which
seven (three) resulted in collisions Mercury-Venus
(Mercury-Sun) (Table~\ref{TabDtEcc}, Fig.~\ref{FigEmax}). 
Thus, the odds for a large increase in Mercury's 
eccentricity found here (0.6\%) are less than 
previous estimates of $\sim$1\% \citep{laskar09}.
\tcx{Strictly, the latter odds were actually $20/2501 = 
0.8$\%. Using different statistical methods 
\citep[e.g.][]{agresti98}, the 95\% 
confidence interval for the two results (0.6\% and
0.8\%) may be estimated as $\pm0.42$\% ($N = 1,600$) and 
$\pm0.36$\% ($N = 2,501$).
Thus at the 95\% confidence level, one can conclude that 
the two results are not statistically different from 
one another, which could be remedied by performing ensemble runs 
with even larger $N$. It is important to realize, however, 
that in order to reduce the 95\% confidence interval for 
the same distributions to, say $\pm0.1$\% would require 
$N \gtrsim 20,000$.}
Note that the 1,600 solutions used initial conditions 
that differed only by 1.75~mm in Mercury's initial radial 
distance between every two adjacent orbits. Yet
the small offsets in Mercury's initial position randomized 
the initial conditions and led to complete divergence of 
trajectories after $\sim$100~Myr (Fig.~\ref{FigEmax}).

\ifTWO \onecolumn \fi

\begin{figure}[p]
\def\epsfsize#1#2{0.35#1}
\vspace*{-12mm}
\hspace*{+1cm}
\centerline{\vbox{
\epsfbox{\figdir 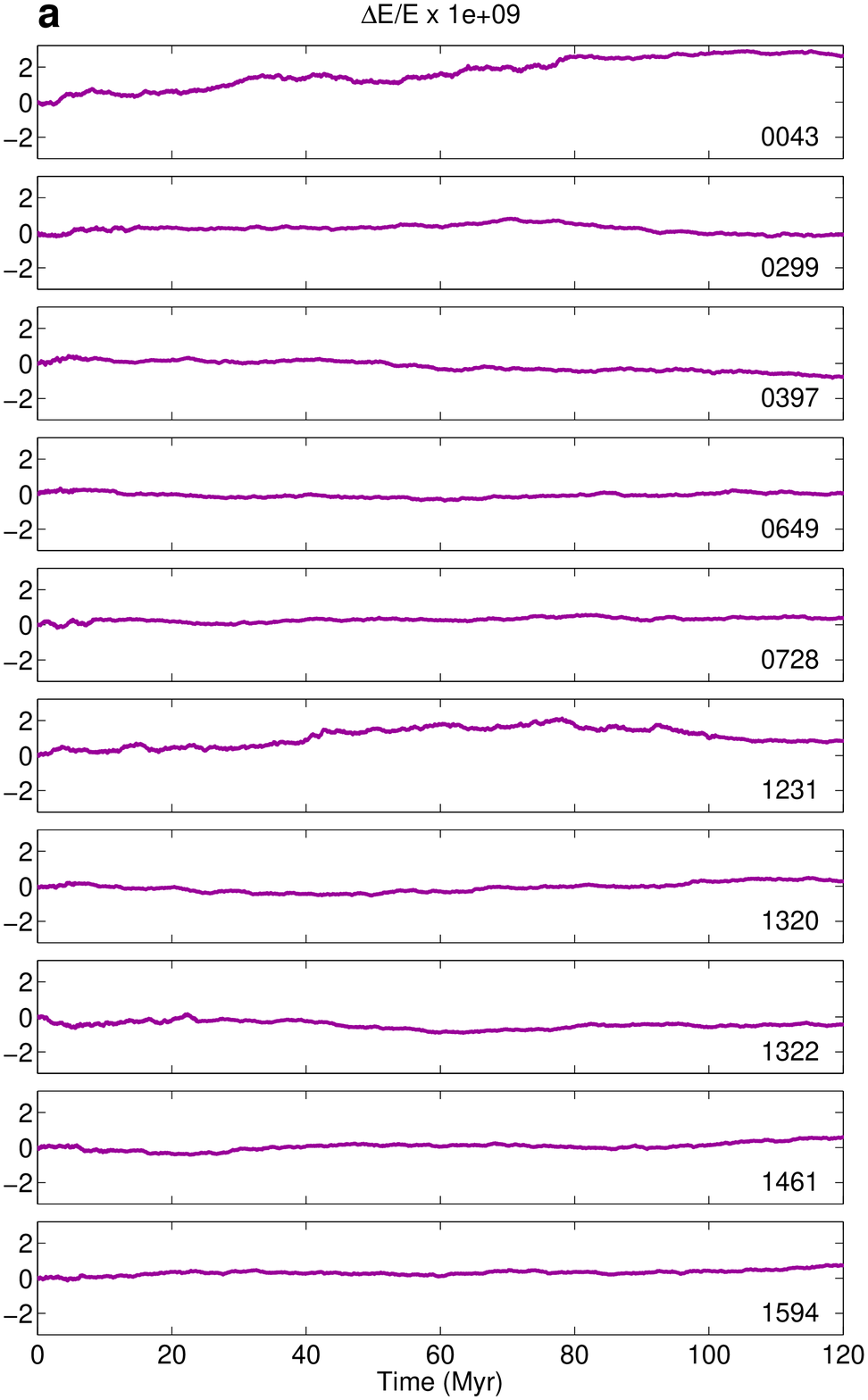}
\epsfbox{\figdir 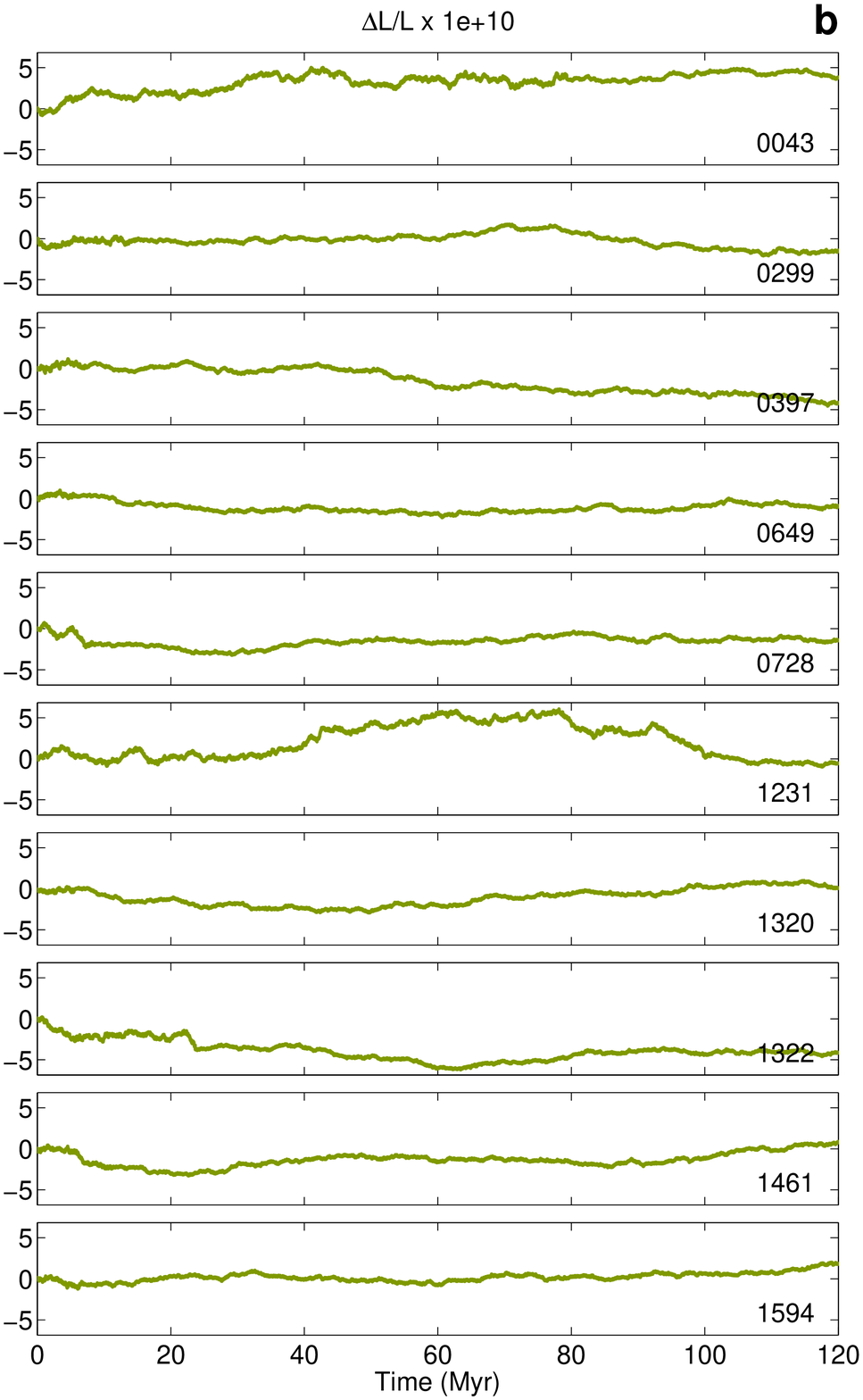}
}}
\hspace*{+1cm}
\centerline{\vbox{
\epsfbox{\figdir 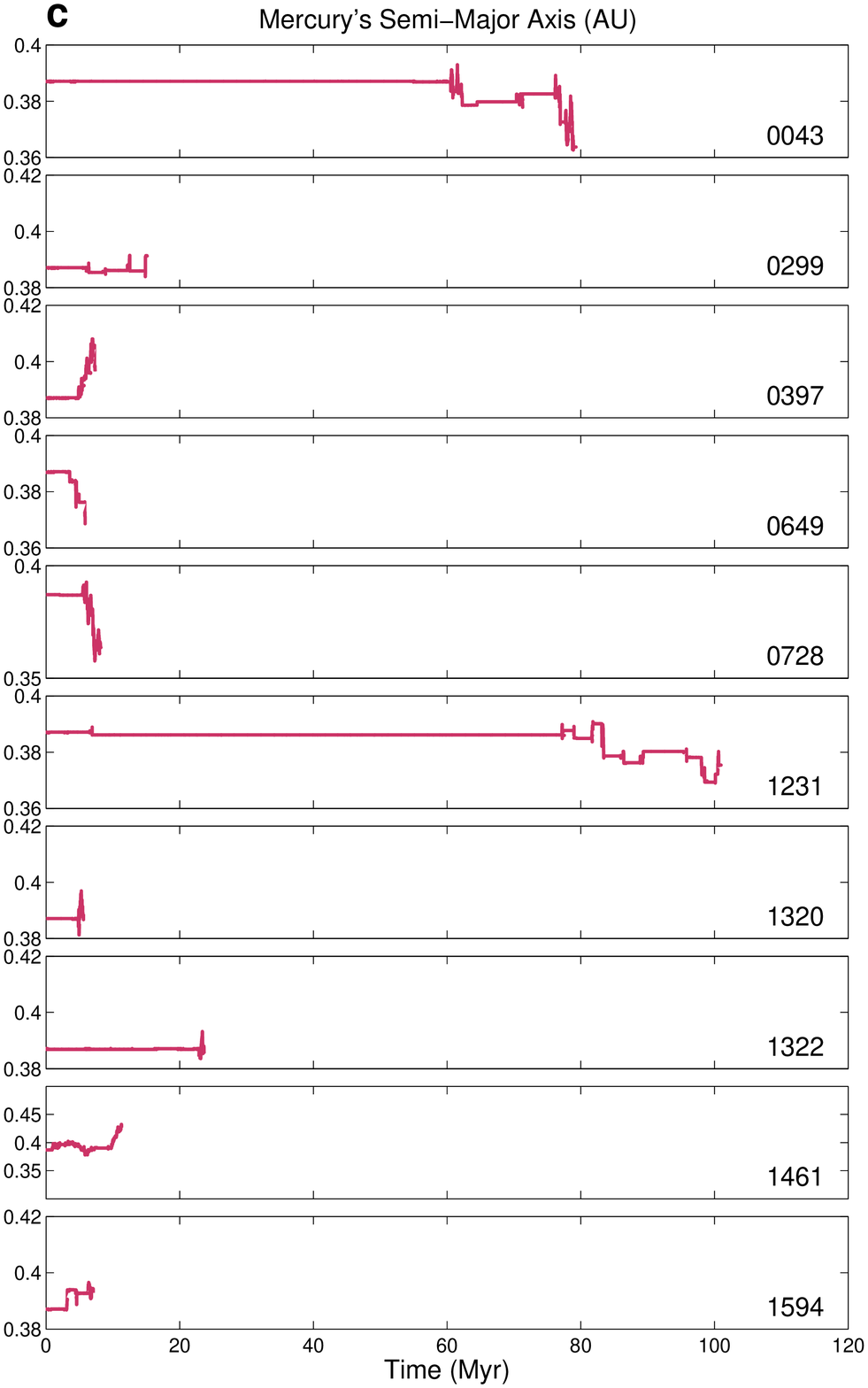}
\epsfbox{\figdir 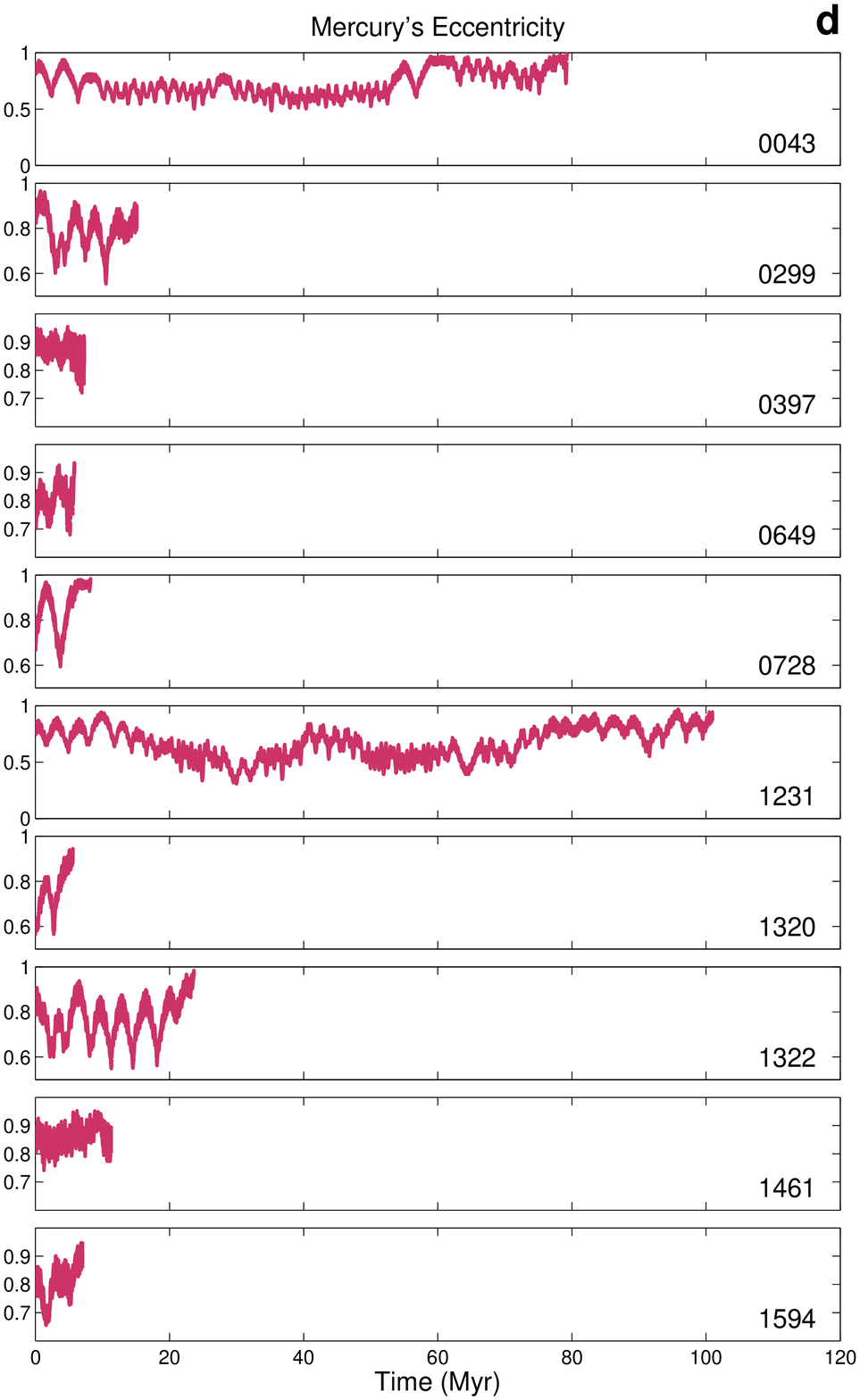}
}}
\caption[]{\small 
120 Myr of 
\BS\ integration of ten solutions, initiated when
oscillations in $|\D E/E|$ and/or shifts in \aM\ 
occurred in the symplectic integrations 
($\eM\gtrsim0.8$, see text). 
(a) Relative error in energy and (b) angular 
momentum. (c) Mercury's semi-major axis and (d)
eccentricity. Note solutions R0043 and R1231.
}
\label{FigBSm6}
\end{figure}

\begin{figure}[t]
\def\epsfsize#1#2{0.38#1}
\vspace*{-12mm}
\hspace*{-8mm}
\centerline{\vbox{
\epsfbox{\figdir 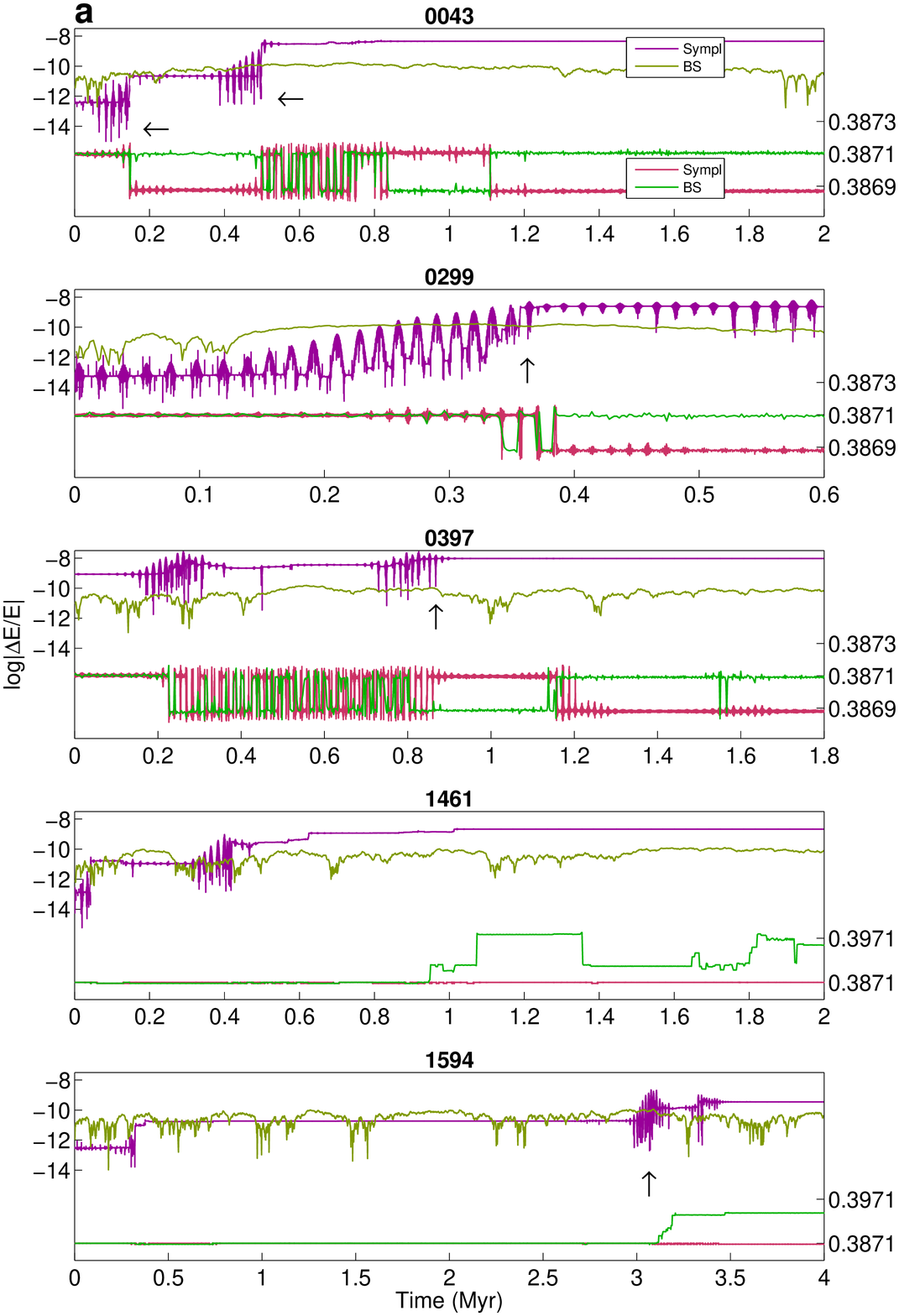}
\epsfbox{\figdir 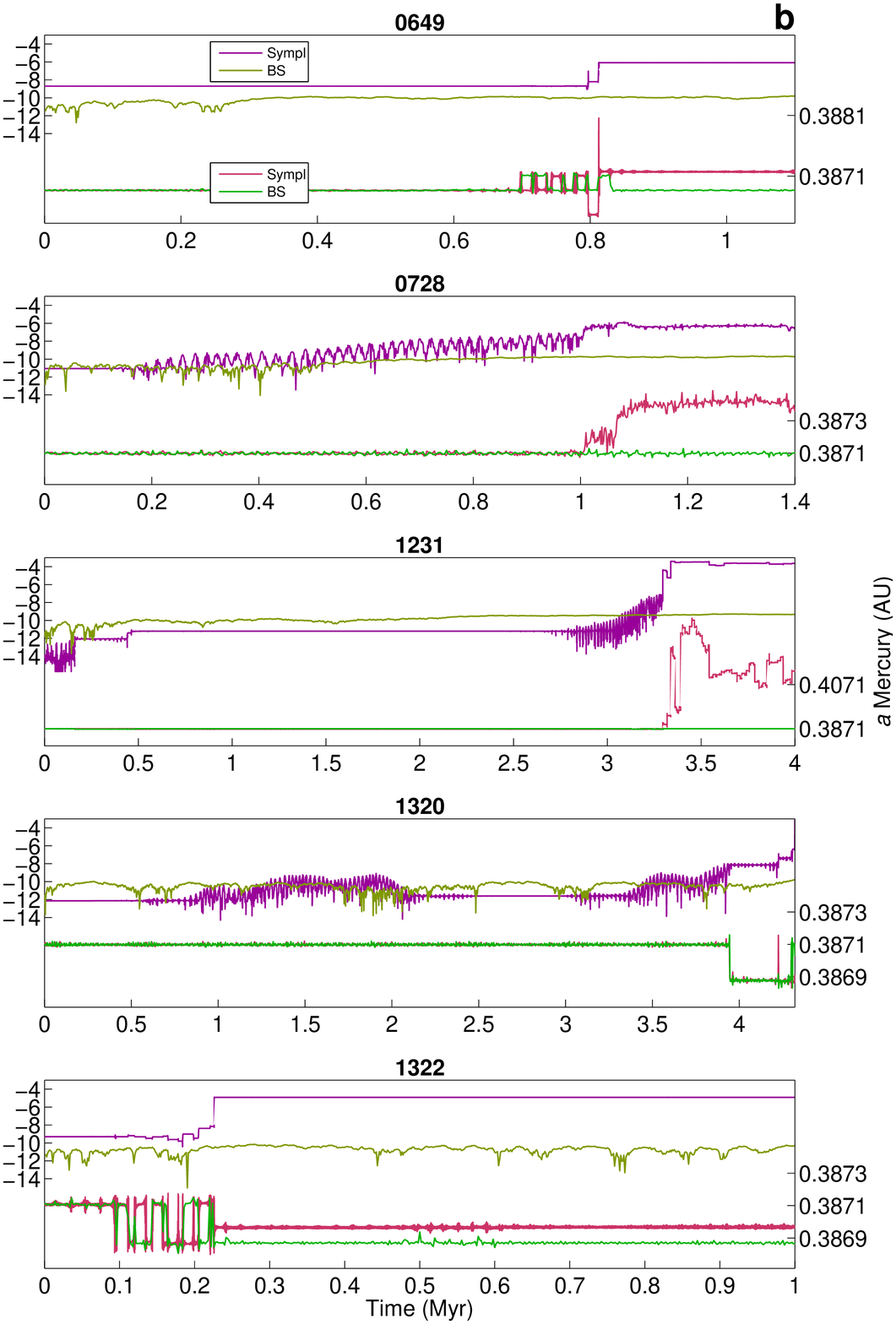}
}}
\caption[]{\small \tcx{
Comparison of symplectic vs.\ BS integration of all ten 
high-\eM\ solutions during the initial phase
(cf.\ Fig.~\ref{FigFail}).
In each panel, the top two graphs 
show $\log{|\D E/E|}$ (left axes), the bottom two 
graphs show \aM\ (right axes). (a) Column of five 
runs with 'initial' symplectic $|\D E/E| \lesssim 10^{-8}$
(see text). Note symplectic $|\D E/E|$ oscillations 
$<10^{-8}$ (arrows, corresponding to high \eM\ 
periods) after which symplectic and BS trajectories 
diverge (indicated by different subsequent \aM\ 
evolutions). (b) Five runs with 'initial' symplectic 
$|\D E/E| > 10^{-8}$ (see text).
}
}
\label{FigFailAll}
\end{figure}

\ifTWO \twocolumn \else \clearpage \fi

\tcx{The results of the symplectic algorithm were used only 
until shifts in \aM\ and/or large $|\D E/E|$ variations 
occurred in the $1/4$~day--symplectic integration 
($\eM > 0.8$, Fig.~\ref{FigEmax}).}
Subsequently, the integration was continued with 
\ms's \BS\ (BS) algorithm (Fig.~\ref{FigBSm6})
(all integrations included GR contributions,
see \info).
As is well known, long-term conservation of energy and 
angular momentum in BS integrations (here: $|\D E/E| 
\lesssim 2\e{-9}$, $|\D L/L| \lesssim 5\e{-10}$) is
generally worse than in symplectic integrations.
However, as mentioned above, for highly eccentric orbits 
the sympletic method can become unstable and may introduce 
artificial chaos \citep{rauch99}. Thus, in the present case
(critical time interval $\lesssim100$~Myr),
it was imperative to integrate high \eM\ solutions
and close encounters with \ms's BS algorithm 
(adaptive stepsize control), specifically 
designed for these tasks \citep{chambers99}.
The symplectic algorithm produced spurious 
results in these situations \tcx{(Figs.~\ref{FigFail},
\ref{FigFailAll}).}

All ten solutions with $\eM > 0.8$ integrated with \BS\
ultimately resulted in collisions involving Mercury; 
Mercury-Venus: R0299, R0397, R0649, R1231, 
R1320, R1461, and R1594; Mercury-Sun: R0043, R0728, and R1322.
Strikingly, in two solutions (R0043 and R1231), 
Mercury continued on highly 
eccentric orbits (after reaching \eM\ values $>$0.93) for 
80-100~Myr before colliding with Venus or the Sun
(Fig.~\ref{FigBSm6}). 
This suggests the potential existence of 
non-catastrophic trajectories, despite Mercury achieving
large eccentricities (is there a chance of recovery,
evading detrimental consequences for the Solar System?).
In contrast, once \eM\ crossed 0.90-0.95 in the symplectic 
integrations ($\Dt = 1/4$~day), large shifts in \aM\ 
and/or energy occurred rapidly, often leading to 
complete destabilization of Mercury's orbit within 
less than \tcx{a few million years (Figs.~\ref{FigFail},
\ref{FigFailAll}).}
Once Mercury was removed (merged with Venus/Sun
via inelastic collision), 
the system behaved very stable (no indication of chaotic 
behavior in the BS integrations).

\subsection{High-\eM\ solutions: Symplectic vs.\ \BS\ results}

\tcx{
As mentioned above,
the symplectic integrations of the ten high-\eM\ solutions 
(smallest $\Dt = 1/4$~day) quickly failed 
once \eM\ reached 0.90-0.95.
In five out of ten runs, the deterioration of the 
integrations was immediately obvious because the 
relative energy error, $|\D E/E|$, rapidly grew 
beyond $10^{-8}$ due to the constant 1/4-day
timestep, which was clearly too large for these
runs (Fig.~\ref{FigFailAll}). However, the symplectic 
$|\D E/E|$ remained 'initially' below $\sim$$10^{-8}$ in the
other five runs, which, if focusing only on the maximum
energy error, may not raise a red flag. Yet, large 
oscillations and jumps in the symplectic $|\D E/E|$
occurred in these five runs during high-\eM\ periods, 
after which the symplectic trajectories rapidly diverged 
from the BS trajectories (Fig.~\ref{FigFailAll}). 
('Initially' here refers to the time interval 
when $|\D E/E| < 10^{-8}$ but trajectories already
diverge; if symplectic \Dt\ is reduced subsequently, 
spurious behavior may go unnoticed). Note that while 
$|\D E/E|$ was still $< 10^{-8}$ after these events, 
$|\D E/E|$ in the symplectic integrations was typically 
two orders of magnitude larger than in the BS integrations
at that point. In runs 0043, 0299, and 0397, \aM\ values 
in the symplectic integrations subsequently dropped below 
those of the BS runs. In runs 1461 and 1594, \aM\ shifted 
to larger values in the BS integrations, which are missed 
in the symplectic integrations. These observations 
reiterate the point made above that even
though the symplectic relative energy error may
remain below a certain threshold (say $<10^{-8}$),
this does not guarantee accurate orbit integration,
for instance, during close encounters and for highly 
eccentric orbits.
}

\tcx{
Ultimately, eight of the ten symplectic high-\eM\ 
integrations resulted in large oscillations and jumps 
in $|\D E/E|$ to values $>$$10^{-8}$. In addition, 
runs 0043, 
0299, and 0728 resulted in rapid destabilization 
of Mercury's orbit ($\eM > 0.99$ and large shifts
in \aM). Two runs (1461 and 1594) conserved $|\D E/E|$
values to below $10^{-8}$ but featured a rapid, 
suspicious decline in \eM\ and missed the \aM\ shifts 
seen in the BS runs (see above). Thus
all ten symplectic high-\eM\ integrations essentially
failed within only $\sim$4~Myr once \eM\ had reached 
critical values between 0.90 and 0.95. The short
lifetime of the highly eccentric Mercurian orbits in the
symplectic integrations (compared to the much longer 
lifetime in the BS integrations, see Fig.~\ref{FigBSm6}) 
is of course mostly a result of too large a timestep, 
which was constant throughout the symplectic integration. 
However, the critical 
point is that even if the symplectic timestep was 
reduced during the integration to maintain a certain 
energy error (\Dt\ can not be changed too often though),
symplectic integrators can easily produce spurious
results for highly eccentric orbits and during close 
encounters, as demonstrated by the comparison with 
the BS integrations.
}

\ifTWO 
\begin{figure}[t]
\twocolumn[{
\def\epsfsize#1#2{0.31#1}
\vspace*{0mm}
\hspace*{3.0cm}
\epsfbox{\figdir 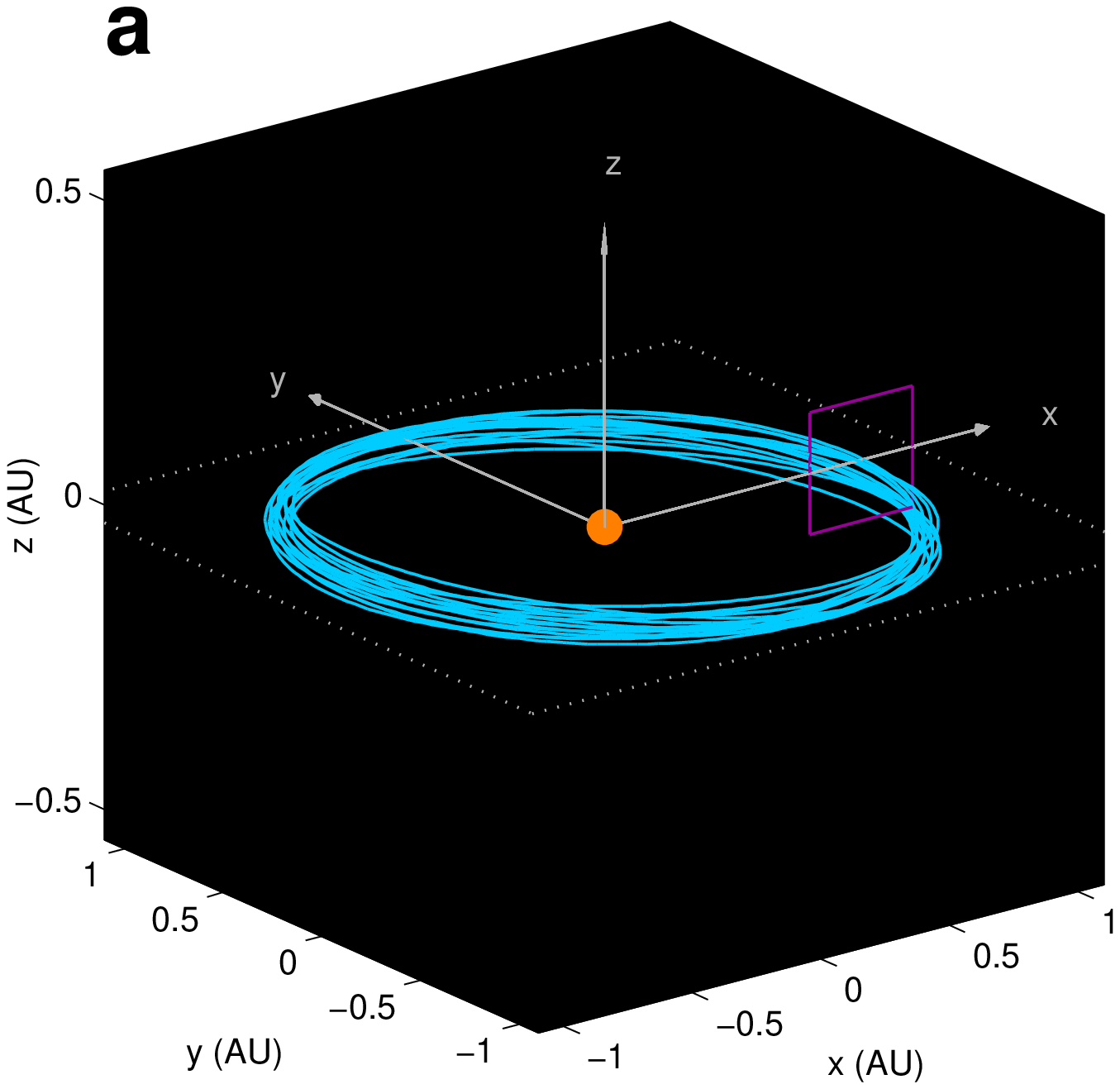}     \hspace*{+3mm}
\epsfbox{\figdir 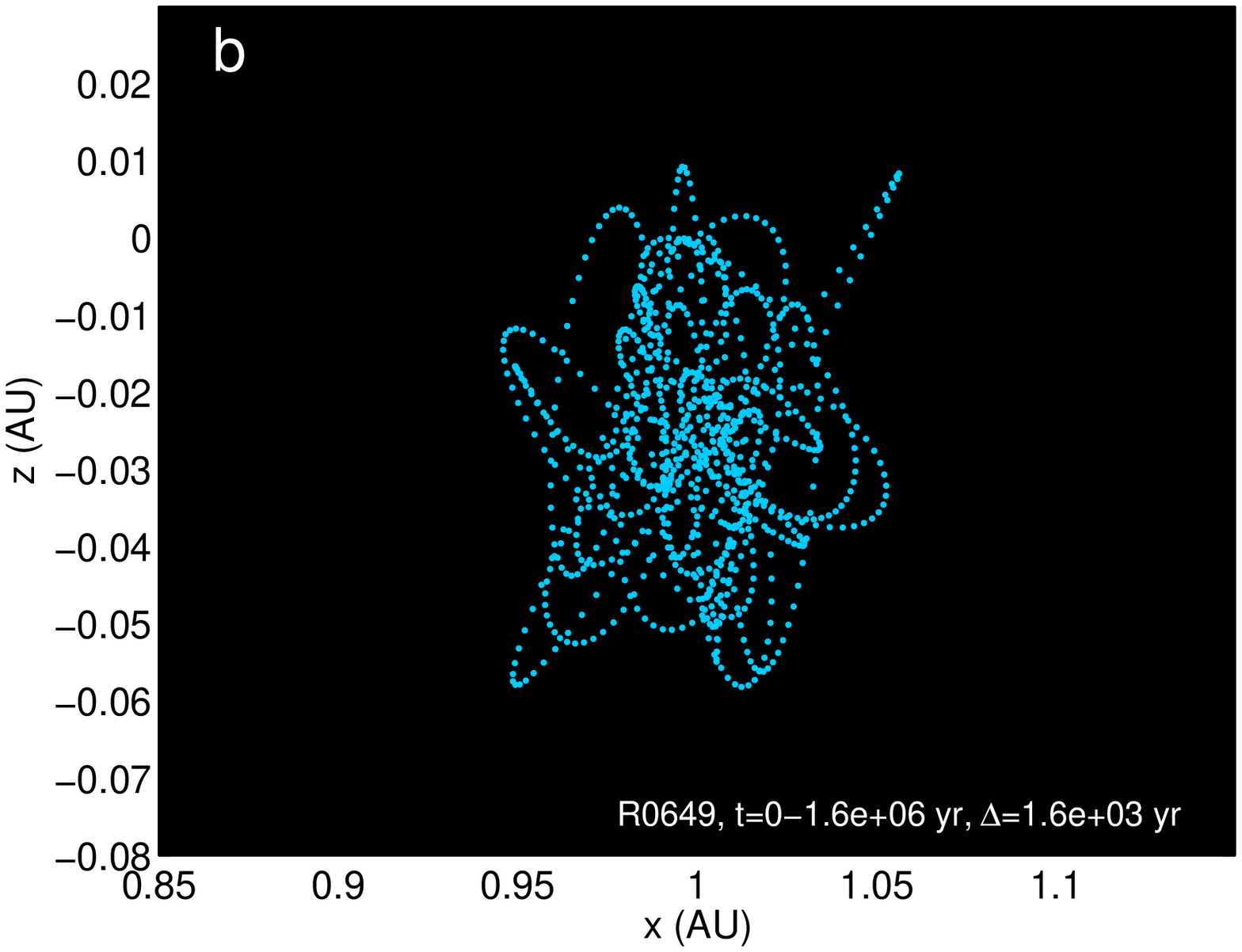} \\
\hspace*{2.0cm}
\epsfbox{\figdir 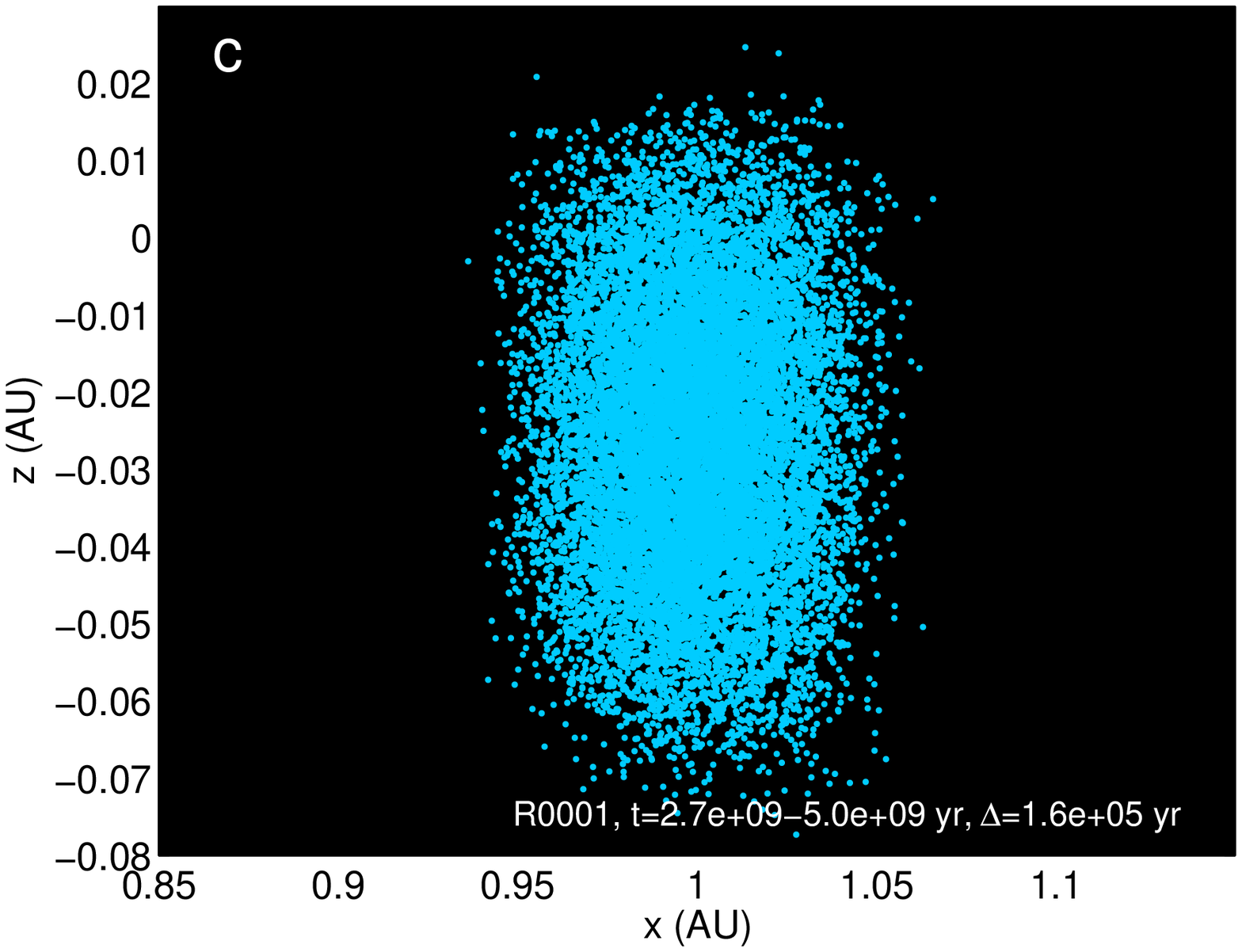}
\epsfbox{\figdir 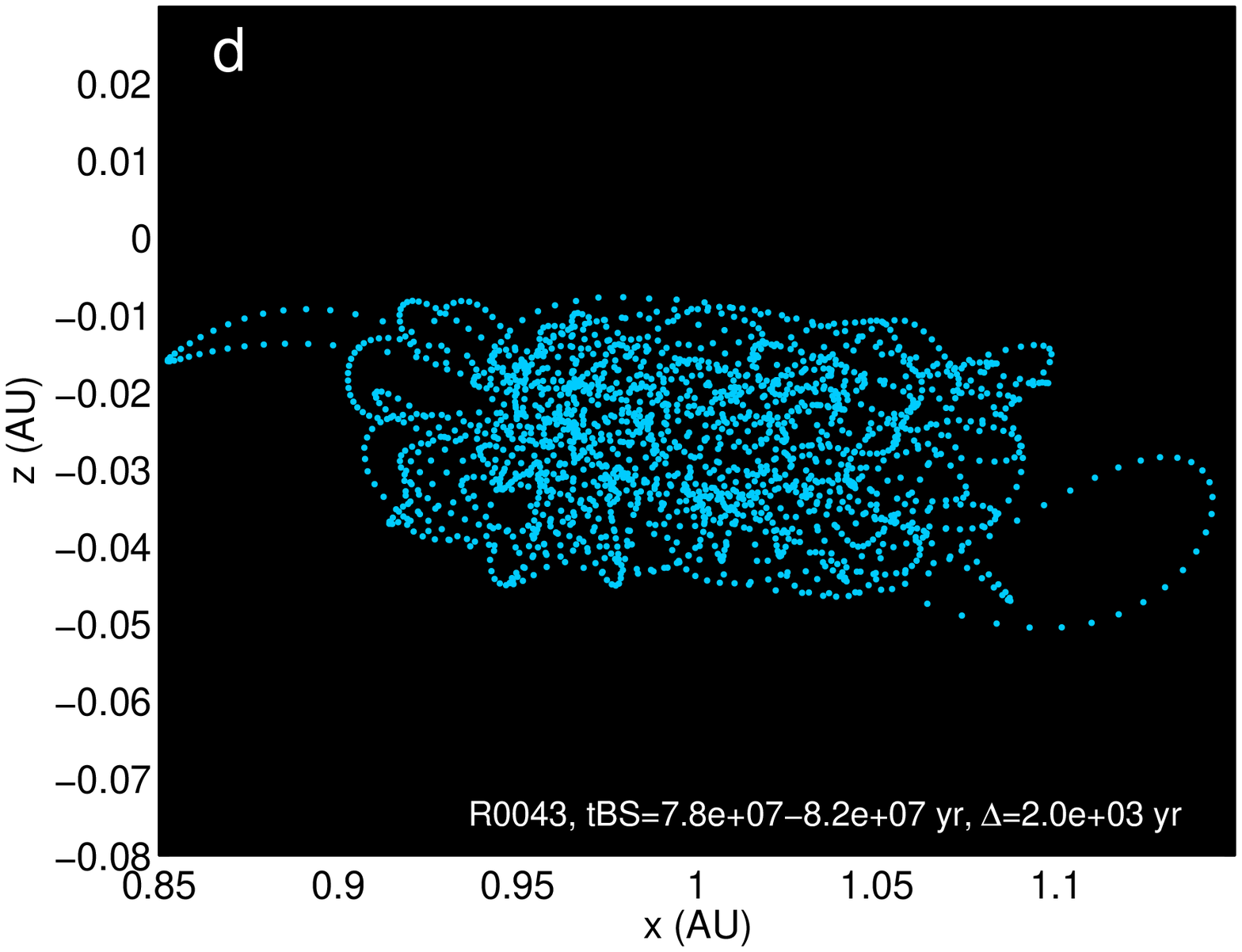} \\
\hspace*{2.0cm}
\epsfbox{\figdir 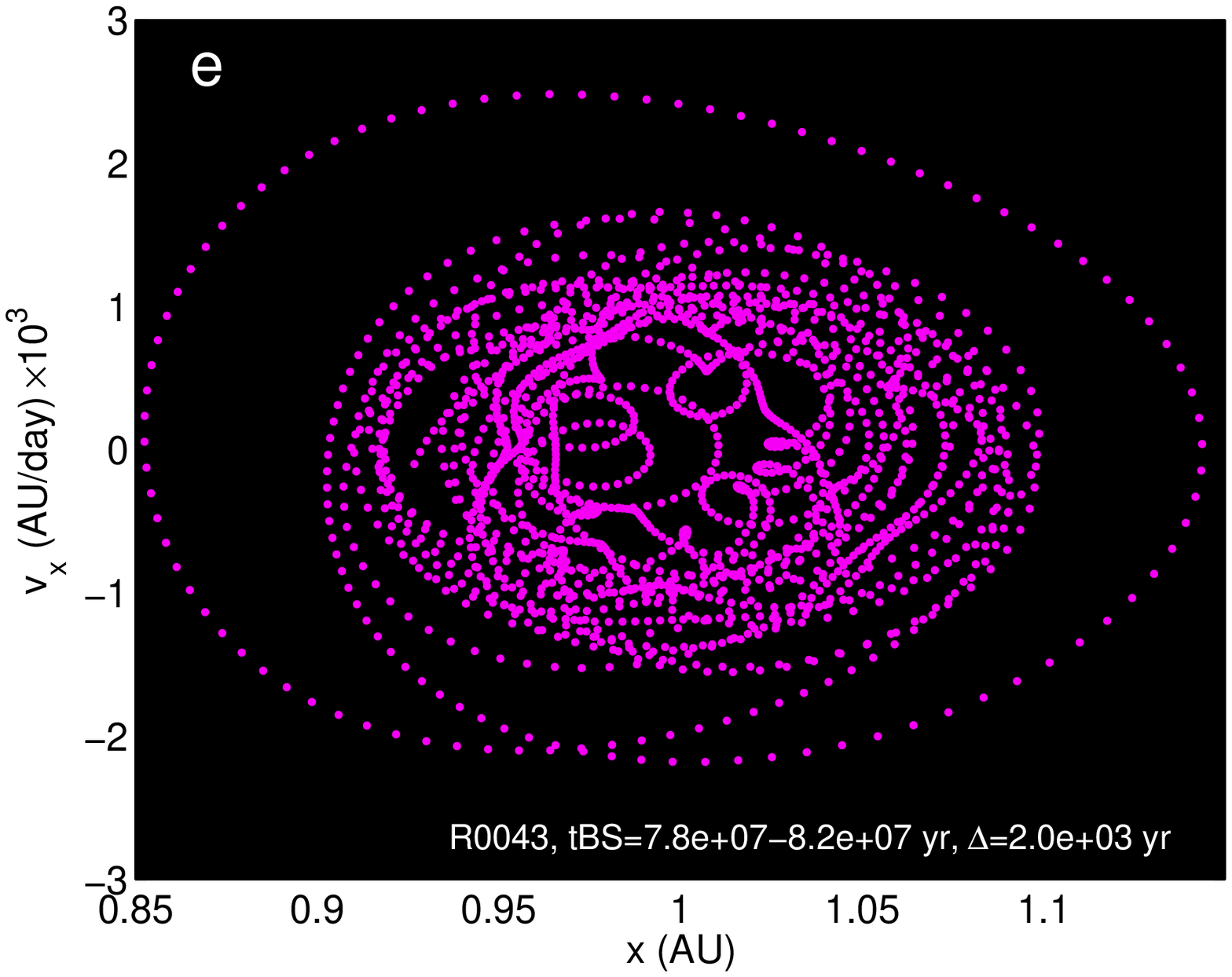} \hspace*{+1mm}
\epsfbox{\figdir 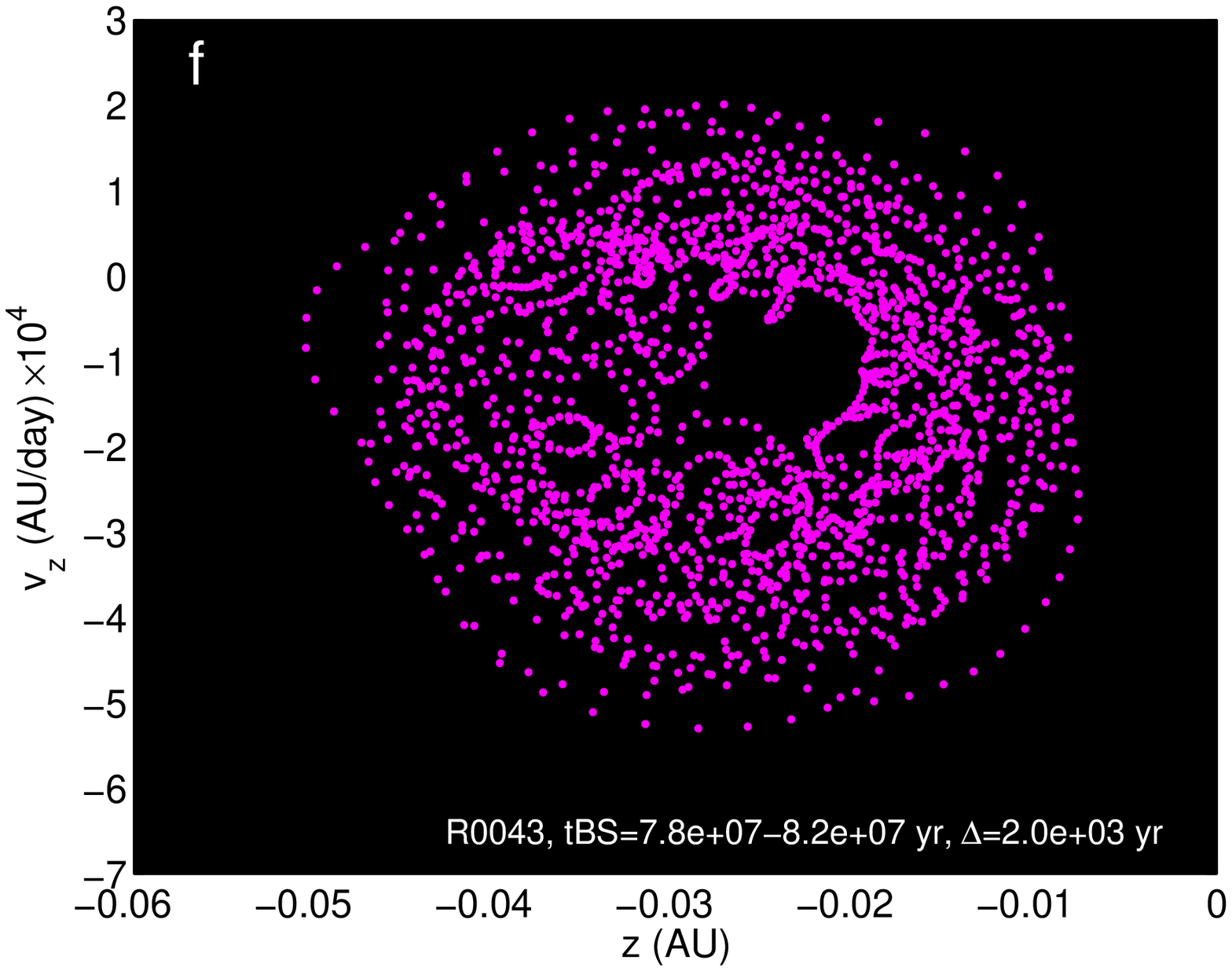}
\caption[]{\small Illustration of the stability of Earth's 
orbit. (a) Earth's orbital path in the heliocentric Cartesian 
system (shown are 15 orbits of R0001, separated by 100~kyr
each). The area of intercept of Earth's trajectory 
with the $xz$-plane ($x>0$) is indicated by the
purple rectangle (size not to scale). (b) Dots
represent the intercepts during 
the first 1.6~Myr of \RSY\ plotted every 1,600~yr.
(c) Intercepts during the final 2.3~Gyr of R0001 plotted 
every 160~kyr.
(d) Intercepts during a 4-Myr period of \RXZ\ plotted every 
2,000~yr. The minimum and maximum $x$-values ($\sim$0.85
and $\sim$1.15~AU)
occur during a $\sim$200~kyr 
interval just before Mercury collides with Venus.
(e) Poincar{\'e} section of Earth's trajectory 
in phase space: $x$-velocity vs.\ $x$-coordinate
when Earth's trajectory crosses the $xz$-plane ($x>0$), 
corresponding to (d). (f) Same as (e) but $z$-velocity vs.\ 
$z$-coordinate.
Note that in all 1,600 solutions integrated here, Earth's 
orbital path intersected the $xz$-plane within an area 
bounded by $x$ and $z$-coordinates that varied at most by 
$\pm$0.15 and $\pm$0.05~AU from the mean. The corresponding 
$x$ and $z$-velocities varied at most by $\pm$2.5\e{-3} 
and $\pm$9.0\e{-4}~AU~day\pmo\ from the mean.
}
}]
\label{FigEOrbit}
\end{figure}
\else  
\begin{figure}[t]
\def\epsfsize#1#2{0.31#1}
\vspace*{-5mm}
\begin{center}
\centerline{\vbox{
\hspace*{+2mm}
\epsfbox{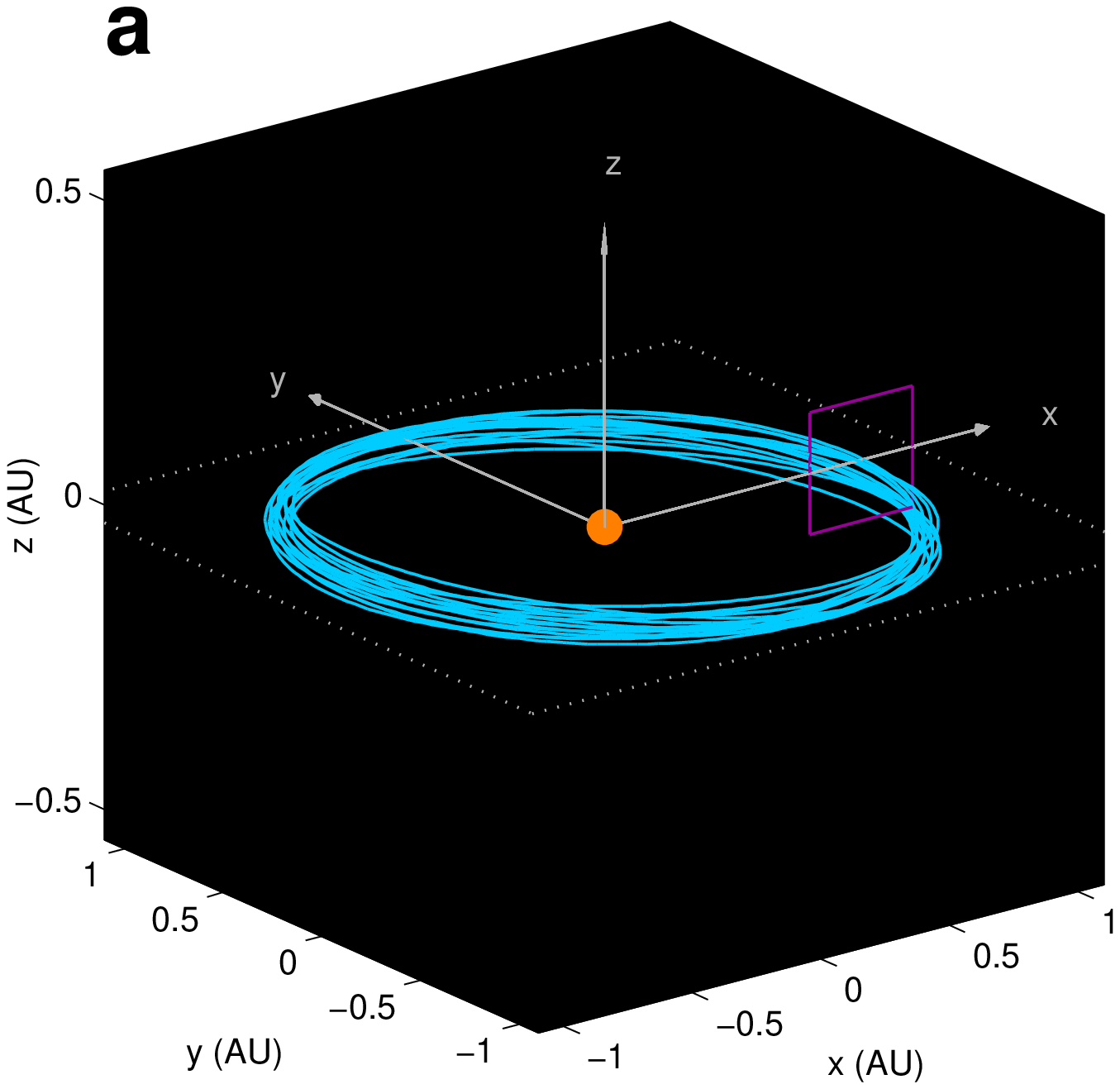}     \hspace*{+5mm}
\epsfbox{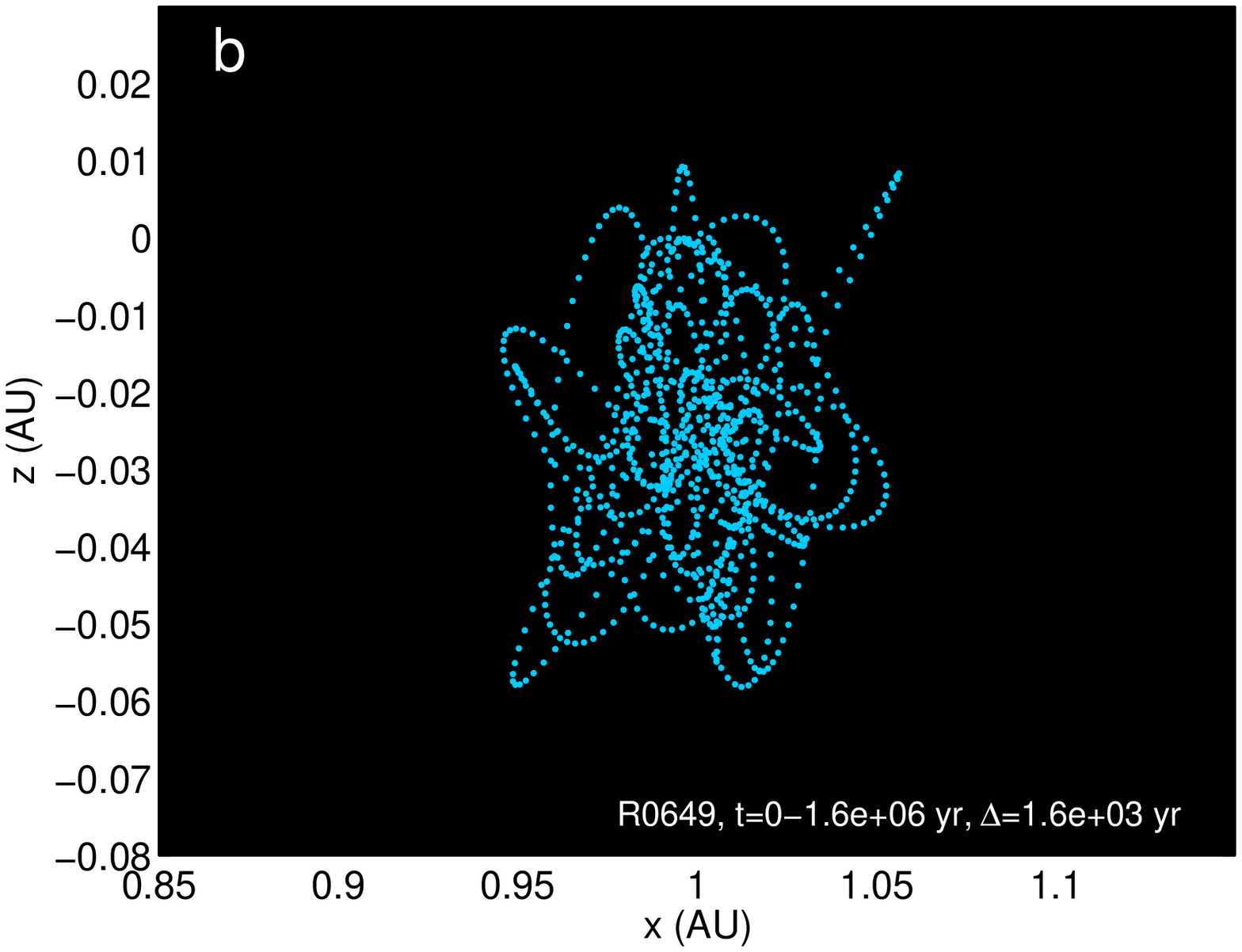} \\[3ex]
\hspace*{-8mm}
\epsfbox{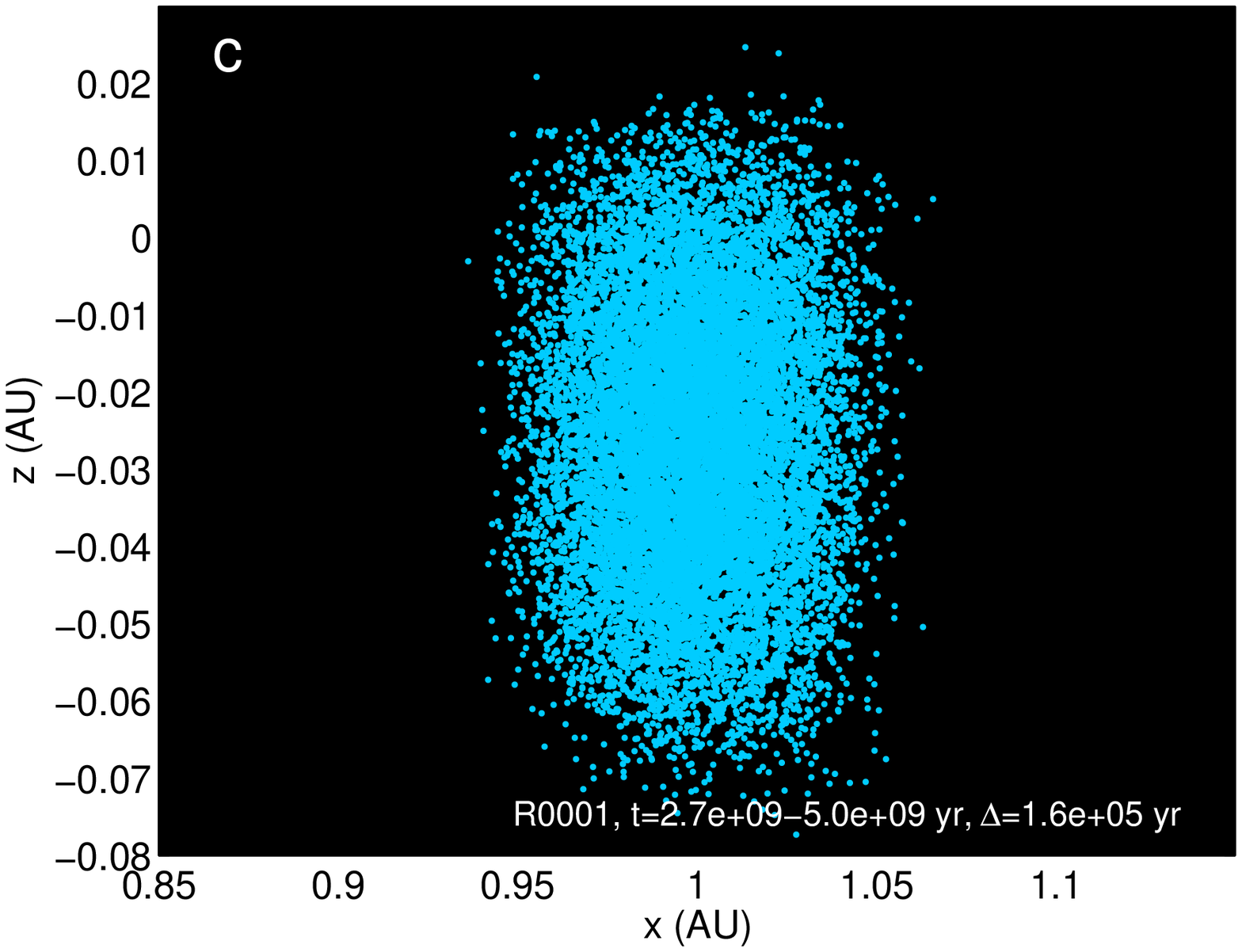}
\epsfbox{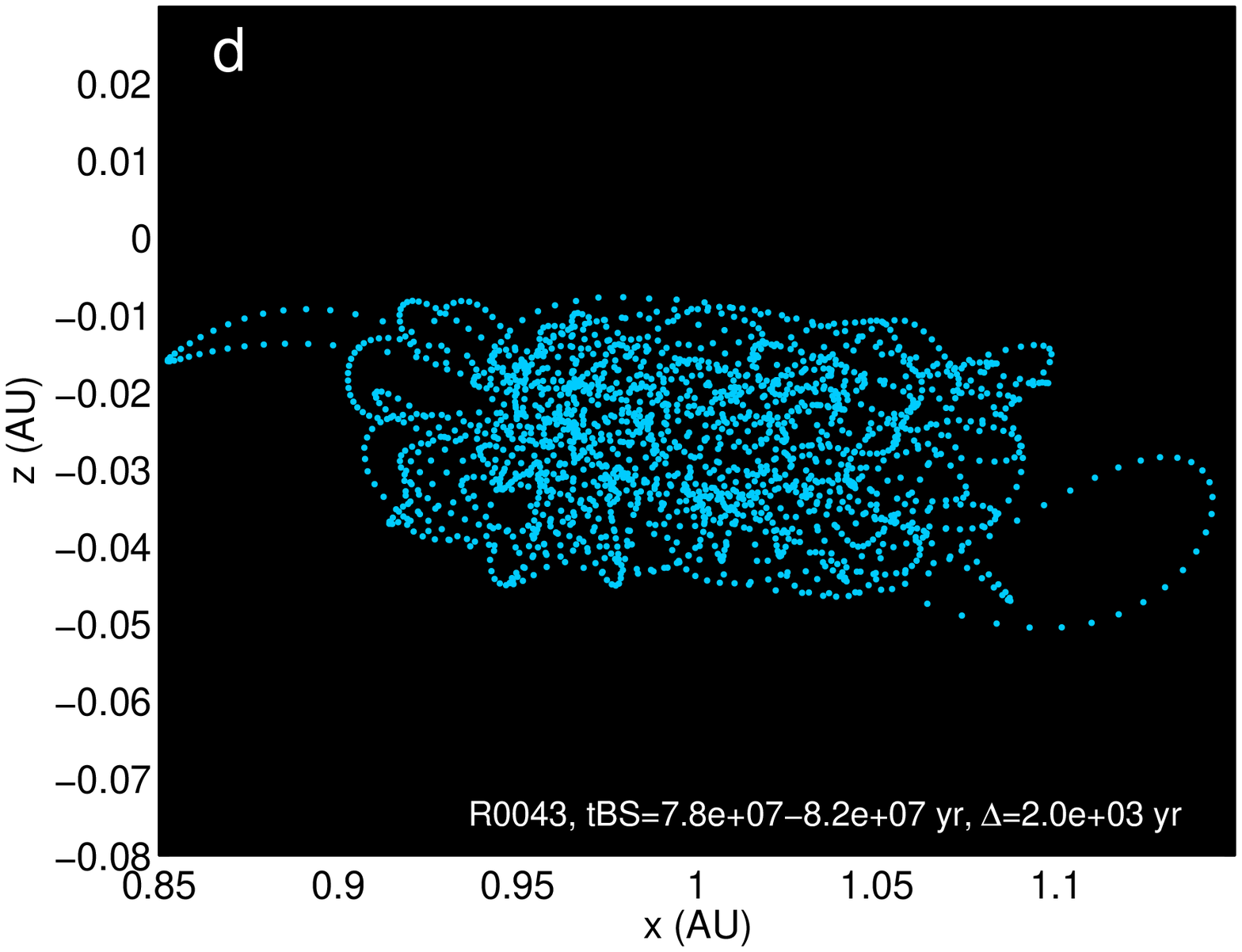} \\
\hspace*{-8mm}
\epsfbox{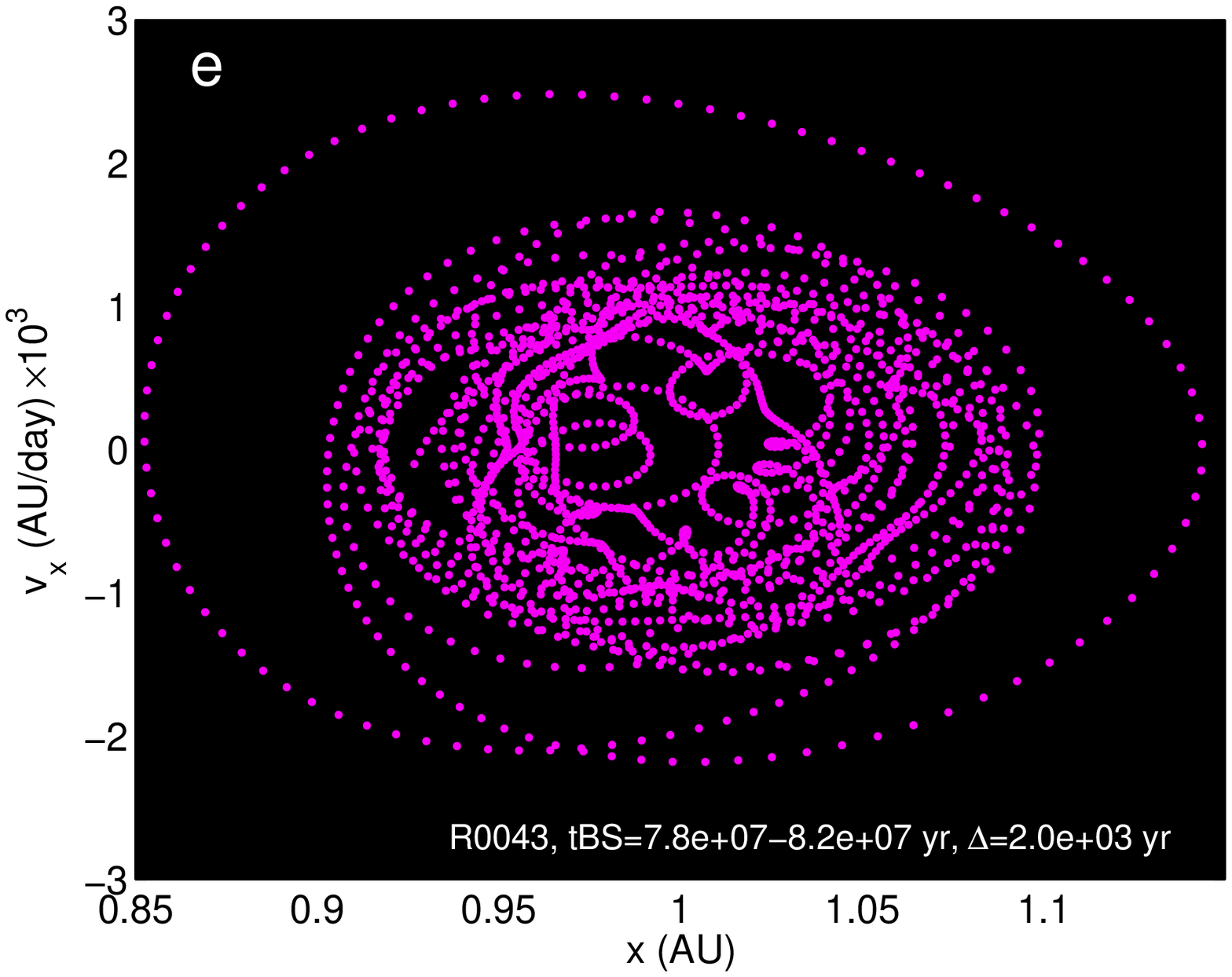} \hspace*{+1mm}
\epsfbox{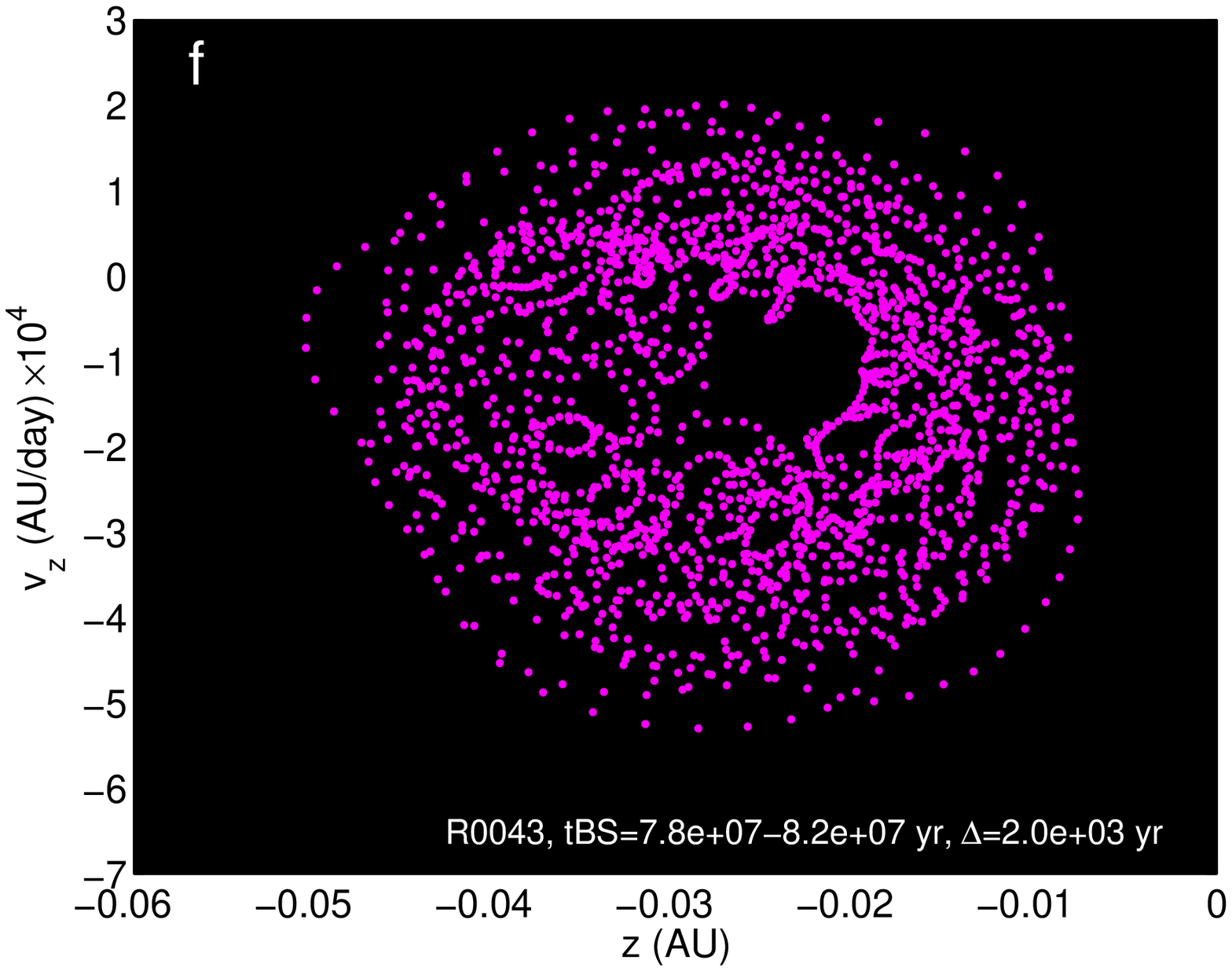}
}}
\end{center}
\caption[]{\small Illustration of the stability of Earth's 
orbit. (a) Earth's orbital path in the heliocentric Cartesian 
system (shown are 15 orbits of R0001, separated by 100~kyr
each). The area of intercept of Earth's trajectory 
with the $xz$-plane ($x>0$) is indicated by the
purple rectangle (size not to scale). (b) Dots
represent the intercepts during 
the first 1.6~Myr of \RSY\ plotted every 1,600~yr.
(c) Intercepts during the final 2.3~Gyr of R0001 plotted 
every 160~kyr.
(d) Intercepts during a 4-Myr period of \RXZ\ plotted every 
2,000~yr. The minimum and maximum $x$-values ($\sim$0.85
and $\sim$1.15~AU)
occur during a $\sim$200~kyr 
interval just before Mercury collides with Venus.
(e) Poincar{\'e} section of Earth's trajectory 
in phase space: $x$-velocity vs.\ $x$-coordinate
when Earth's trajectory crosses the $xz$-plane ($x>0$), 
corresponding to (d). (f) Same as (e) but $z$-velocity vs.\ 
$z$-coordinate.
Note that in all 1,600 solutions integrated here, Earth's 
orbital path intersected the $xz$-plane within an area 
bounded by $x$ and $z$-coordinates that varied at most by 
$\pm$0.15 and $\pm$0.05~AU from the mean. The corresponding 
$x$ and $z$-velocities varied at most by $\pm$2.5\e{-3} 
and $\pm$9.0\e{-4}~AU~day\pmo\ from the mean.
}
\label{FigEOrbit}
\end{figure}
\fi 

\subsection{Future evolution of Earth's orbit}

Mercury's orbital dynamics had little effect on 
Earth's orbit as none of the 1,600 solutions showed 
a destabilization of Earth's future orbit over the 
next 5~Gyr. Rather, 
Earth's future orbit was highly stable.
For illustration, during a typical, full 5-Gyr run, 
Earth's orbital path typically intersected the $xz$-plane 
of the heliocentric Cartesian system
within an area bounded by $x$ and $z$-coordinates that
varied at most by $\pm$0.07 and $\pm$0.05~AU 
from the mean, respectively (Fig.~\ref{FigEOrbit}).
In the most extreme case found here (\RXZ), $x$ and 
$z$ varied at maximum by $\pm$0.15 and 
$\pm$0.05~AU (Earth's eccentricity approaching
0.15). Importantly, the maximum $x,z$ 
variations in \RXZ\ were actually restricted to a 
$\sim$200~kyr interval just before Mercury collided with 
Venus (Fig.~\ref{FigEOrbit}).
Poincar{\'e} sections of Earth's trajectory 
in phase space showed that the corresponding 
$x$ and $z$-velocities in all 1,600 solutions
varied at most by $\pm$2.5\e{-3} 
and $\pm$9.0\e{-4}~AU~day\pmo\ from the mean
(Fig.~\ref{FigEOrbit}).

Also, none of the 1,600 simulations led to a close 
encounter, let alone a collision, involving the Earth. 
The minimum distance 
between the Earth and another planet (viz.\ Venus, 
$\dm \simeq 0.09$~AU) occurred in \RDE\ during
a 50-year period
when Mercury collided with the Sun. This \dm\ 
does not qualify as a close encounter though; it is still
$\sim$9 times larger than the Earth-Venus mutual 
Hill radius, $\rH \simeq 0.01$~AU \citep{chambers96},
or $\sim$1/3 of the current \dm\ ($\sim$$30 \x \rH$).
A previous study suggested the possibility of a 
collision between the Earth and Venus via transfer of 
angular momentum from the giant planets to the 
terrestrial planets \citep{laskar09}. However, the
total destabilization of the inner Solar System
only occurred after \eM\ had already crossed 0.9 and 
$|\D E/E|$ had grown beyond $2\e{-8}$ in their 
symplectic integration \citep{laskar09}.
Did such disastrous trajectories for the Earth arise 
because Mercury's periapse and close 
encounters were not adequately resolved at some 
point in the symplectic integration?

\ifTWO
\begin{figure}[t]
\def\epsfsize#1#2{0.47#1}
\hspace*{-8mm}
\else
\begin{figure}[t]
\def\epsfsize#1#2{0.55#1}
\vspace*{-8mm}
\hspace*{+0cm}
\begin{center}
\fi
\centerline{\vbox{\epsfbox{\figdir 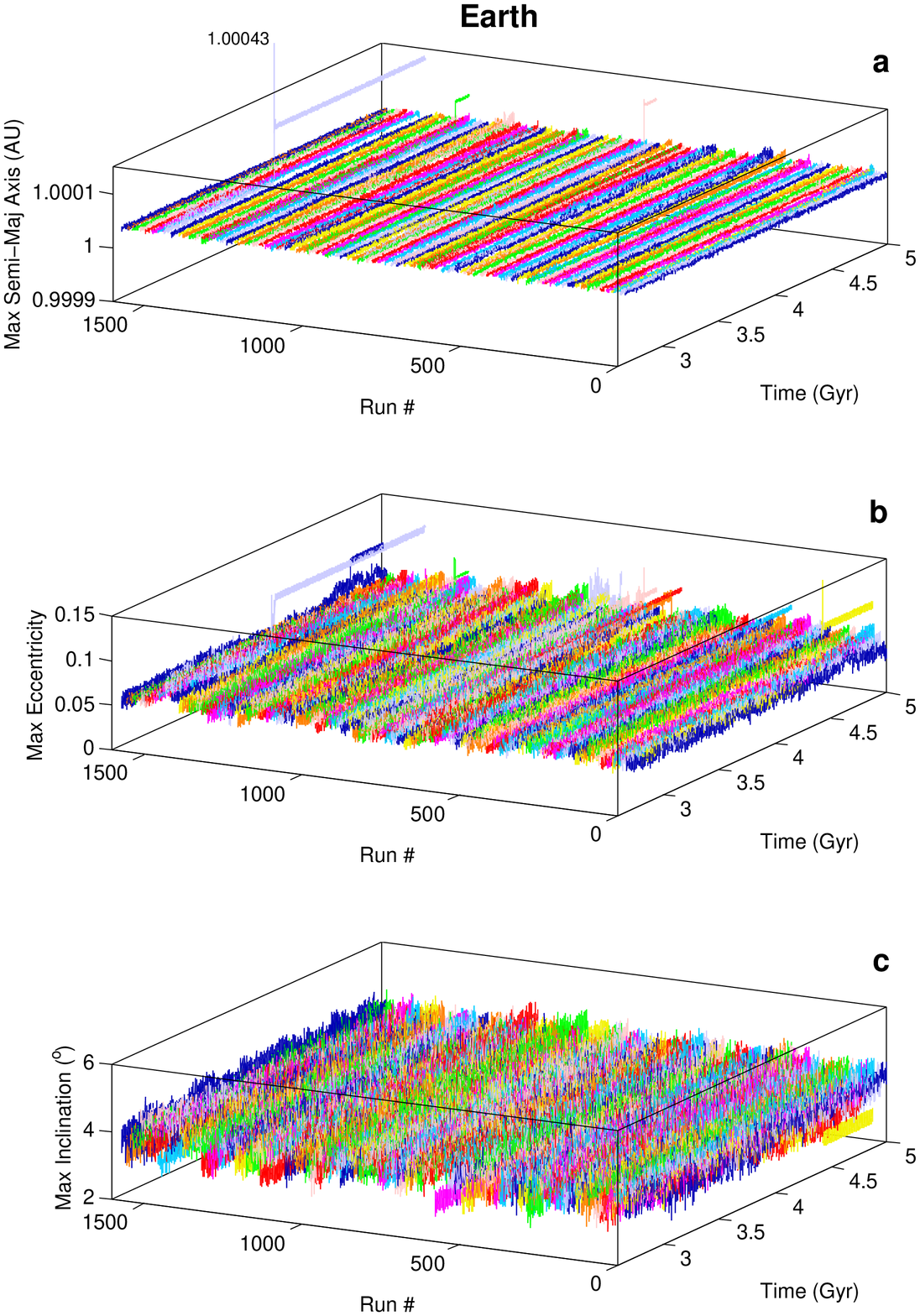}}}
\ifONE \end{center} \fi
\caption[]{\small 
Maximum values per 1-Myr bin of Earth's slowly changing 
orbital elements. (On gyr-time scale, argument of periapsis, 
longitude of ascending node, and mean anomaly may be
considered 'fast angles'.)
For better visualization, only every 20th of the 1,600 solutions 
are plotted plus all runs with $\eM > 0.8$ for $t = 2.5-$5~Gyr.
The 1,600 solutions used initial conditions that 
differed only by 1.75~mm in Mercury's initial radial distance
between every two adjacent orbits.
(a) Maximum semi-major axis, (b) eccentricity, and (c) inclination.
All three elements remained within narrow bands in the 1,600 
solutions: 
$\max(\aE) < 1.0005$~AU,
$\max(\eE) < 0.15$, and
$\max(\iE) < 5.1\deg$.
The runs with elevated \aE\ and \eE\ are solutions with
large increases in Mercury's eccentricity. For example, 
\aE\ reached 1.00043~AU in \RXA\ during a high-\eM\ interval,
which occurred $\sim$8~Myr before Mercury collided
with Venus. After Mercury had been merged with Venus, 
the system behaved very stable.
}
\label{FigEaimax}
\end{figure}

The values of Earth's orbital elements semi-major axis (\aE), 
eccentricity (\eE), and inclination (\iE) remained within 
narrow bands in all 1,600 solutions studied here 
(Fig.~\ref{FigEaimax}). For instance, 
$\max(\aE) < 1.0005$~AU,
$\max(\eE) < 0.15$, and
$\max(\iE) < 5.1\deg$
(for comparison, J2000 mean values are
$\aE = 1.0000$,
$\eE = 0.017$,
$\iE = 0.0\deg$).
As the present study essentially sampled eight 
trillion annual realizations of possible Solar System 
configurations in the future ($1,600\x5\cdot10^{9}$),
the near constancy of Earth's orbital elements 
$a, e$ and $i$ suggests that Earth's orbit is dynamically 
highly stable for billions of years to come.

\section{Conclusions}

In the present manuscript, I have reported results from 
computationally demanding ensemble integrations 
($N = 1,600$) of the Solar System's full equations 
of motion at unprecedented accuracy over 5~Gyr.
All integrations included relativistic corrections,
which substantially reduce the probability for Mercury's 
orbit to achieve large eccentricities.
The computations show that the relative energy 
error in symplectic integrations (say $\lesssim10^{-8}$) 
is not a sufficient criterion to ensure accurate steps 
for highly eccentric orbits and during close encounters.
The calculated odds for a large increase in 
Mercury's eccentricity are less than previously estimated.
Most importantly, none of the 1,600 solutions led to
a close encounter, let alone a collision, involving the Earth.
I conclude that Earth's orbit is dynamically highly stable
for billions of years in the 
future, despite the chaotic behavior of the Solar System.
A dynamic life-time of $\gtrsim5$~Gyr into the future may 
be somewhat short of the Sun's red giant phase when most
inner planets will likely be engulfed, but clearly
exceeds Earth's estimated future habitability of 
perhaps another 1-3~Gyr \citep{schroeder08,rushby13}.

\vfill
\noindent
{\small
{\bf Acknowledgments.}
This project would not have been
possible without the HPC cluster test-phase of the 
University of Hawaii. I thank Ron Merrill, 
Sean Cleveland, and Gwen Jacobs for providing the 
opportunity to participate in the test program.
The anonymous reviewer is thanked for valuable comments,
which improved the manuscript.
I am grateful to Peter H. Richter who dared to introduce
us to Chaos, Poincar{\'e}, and Solar System dynamics
in a 1989-undergraduate physics course on classical 
mechanics. Nemanja Komar's assistance in analyzing the 
numerical output was greatly appreciated. 
}

\appendix

\section{Contributions from general relativity}

Relativistic corrections \citep{einstein16} are critical 
as they substantially reduce the probability for Mercury's 
orbit to achieve large eccentricities 
\citep{laskar09,zeebe15apj}. General relativity (GR)
corrections are available in \hnb\ but not in \ms.
First  
Post-Newtonian (1PN) corrections \citep{soffel89,poisson14}
were therefore implemented before using \ms's \BS\ 
algorithm.

Because of the Sun's dominant mass, GR effects are
considered only between each planet and the Sun, not
between planets. Relative to the barycenter,
the 1PN acceleration (denoted 
by ${\widetilde{ \ \ }}$\ ) 
due to the Sun's mass
$m_1$ on planet $j = 2,N$ may be
written as \citep{soffel89,poisson14}:
\beqn
\v{\at}_j & = & \q{1}{c^2} \left\{
\q{G m_1}{r_j^3} \left[ 
-v_j^2 - 2v_1^2 + 4( \v{v}_j   \cdot \v{v}_1 )
  + \q{3}{2r^2_j} ( \v{x}^h_j \cdot \v{v}_1 )^2 
  + \q{5Gm_j}{r_j} + \q{4Gm_1}{r_j}
\right] \v{x}^h_j \right. \nn \\
& & + \left.
\q{G m_1}{r_j^3} [\v{x}^h_j \cdot (4 \v{v}_j - 3 \v{v}_1 )]
\ \v{v}^h_j
\right\} \ ,
\eeqn
where $\v{x}$'s and $\v{v}$'s are barycentric 
positions and velocities, the superscript $h$ refers 
to heliocentric coordinates, and 
$r_j = |\v{x}^h|$. The 1PN acceleration
on the Sun is:
\beqn
\v{\at}_1 & = & \sum_{j=2}^N \nn \\ & &
\q{1}{c^2} \left\{
\q{G m_j}{r_j^3} \left[ 
 v_1^2 + 2v_j^2 - 4( \v{v}_j   \cdot \v{v}_1 )
  - \q{3}{2r^2_j} ( \v{x}^h_j \cdot \v{v}_j )^2 
  - \q{4Gm_j}{r_j} - \q{5Gm_1}{r_j}
\right] \v{x}^h_j \right. \nn \\
& & + \left.
\q{G m_j}{r_j^3} [\v{x}^h_j \cdot (4 \v{v}_1 - 3 \v{v}_j )]
\ \v{v}^h_j
\right\} \ .
\eeqn
As \ms's \BS\ algorithm uses heliocentric coordinates
($\v{x}^h_j = \v{x}_j - \v{x}_1$), the 1PN acceleration 
on planet $j$ in heliocentric coordinates
is required, given by:
\beqn
\v{\at}^h_j & = & \v{\at}_j - \v{\at}_1 \ .
\eeqn
The above equations were implemented in the force
calculation routines of \ms.

Furthermore, the energy and angular momentum correction
terms in the 1PN approximation are \citep{poisson14}:
\beqn
\Et_j & = &
\q{\eta_j M_j}{c^2} \left\{
\q{3}{8}(1 - 3\eta_j) (v^h_j)^4 + 
\q{G M_j}{2r_j} \left[ 
    (3+\eta_j) (v^h_j)^2
  + \q{\eta_j}{r_j^2} ( \v{x}^h_j \cdot \v{v}^h_j )^2 
\right] 
  + \q{G^2M_j^2}{2r_j^2}
\right\} \\
\v{\Lt}_j & = & 
\q{\eta_j M_j}{c^2} \left[
\q{1}{2}(1 - 3\eta_j) (v^h_j)^2 + 
(3+\eta_j) \q{G M_j}{r_j}
\right]  \v{x}^h_j \x \v{v}^h_j \ ,
\eeqn
where 
$\eta_j = m_1 m_j/M_j^2$ and $M_j = m_1+m_j$. These
correction
terms were added to the routine \verb|mxx_en()|, which
computes energy and angular momentum in \ms.
Finally, the 1PN Solar System barycenter in the 
present approximation is given by \citep{newhall83}:
\beqn
\v{0} = 
m_1 \v{x}_1 \left( 
1 + \q{1}{2c^2} v_1^2 
- \sum_{j=2}^N  \q{G m_j}{2c^2r_j}
\right)
+ \sum_{j=2}^N
m_j \v{x}_j \left( 
1 + \q{1}{2c^2} v_j^2 - \q{G m_1}{2c^2r_j}
\right) \ ,
\eeqn
which was used in the conversion between heliocentric
and barycentric coordinates.

The results obtained with \ms\ and the above GR 
implementation may be compared to results obtained 
with \hnb\ (both \BS, relative accuracy $10^{-15}$). 
For example, 
over the 21st century, Mercury's average perihelion 
precession (only due to GR) was 0.42977''~y\pmo\ computed with
\hnb\ and 0.42976''~y\pmo\ computed with \ms\ and 
1PN corrections. In terms of Mercury's eccentricity 
(\eM) evolution, the difference in \eM\ between \hnb\ 
and \ms\ runs (both \BS\ with GR correction) was 
$\lesssim10^{-5}$ over 2~Myr starting at present
initial conditions.

\renewcommand{\baselinestretch}{\blsC}\selectfont

%

\begin{thebibliography}{}
\expandafter\ifx\csname natexlab\endcsname\relax\def\natexlab#1{#1}\fi

\bibitem[{Agresti \& Coull(1998)}]{agresti98}
Agresti, A., \& Coull, B.~A. 1998, Am. Stat., 52, 119

\bibitem[{{Batygin} \& {Laughlin}(2008)}]{batygin08}
{Batygin}, K., \& {Laughlin}, G. 2008, Astrophys.~J., 683, 1207

\bibitem[{{Batygin} {et~al.}(2015){Batygin}, {Morbidelli}, \&
  {Holman}}]{batygin15}
{Batygin}, K., {Morbidelli}, A., \& {Holman}, M.~J. 2015, Astrophys.~J., 799,
  120

\bibitem[{{Chambers}(1999)}]{chambers99}
{Chambers}, J.~E. 1999, Mon. Not. R. Astron. Soc., 304, 793

\bibitem[{{Chambers} {et~al.}(1996){Chambers}, {Wetherill}, \&
  {Boss}}]{chambers96}
{Chambers}, J.~E., {Wetherill}, G.~W., \& {Boss}, A.~P. 1996, Icarus, 119, 261

\bibitem[{{Einstein}(1916)}]{einstein16}
{Einstein}, A. 1916, Annalen der Physik, VI. Folge, 49(7), 769

\bibitem[{{Ito} \& {Tanikawa}(2002)}]{ito02}
{Ito}, T., \& {Tanikawa}, K. 2002, \mnras, 336, 483

\bibitem[{Laplace(1951)}]{laplace51}
Laplace, P.~S. 1951, in {} (Dover Publications, New York)

\bibitem[{{Laskar}(1989)}]{laskar89}
{Laskar}, J. 1989, Nature, 338, 237

\bibitem[{Laskar(2013)}]{laskar13}
Laskar, J. 2013, in Progress in Mathematical Physics, Vol.~66, Chaos, ed.
  B.~Duplantier, S.~Nonnenmacher, \& V.~Rivasseau (Springer Basel), 239--270

\bibitem[{{Laskar} \& {Gastineau}(2009)}]{laskar09}
{Laskar}, J., \& {Gastineau}, M. 2009, Nature, 459, 817

\bibitem[{{Morbidelli}(2002)}]{morbidelli02}
{Morbidelli}, A. 2002, {Modern Celestial Mechanics: Aspects of Solar System
  Dynamics} (Taylor {\&} Francis, London)

\bibitem[{{Murray} \& {Holman}(1999)}]{murray99}
{Murray}, N., \& {Holman}, M. 1999, Science, 283, 1877

\bibitem[{{Newhall} {et~al.}(1983){Newhall}, {Standish}, \&
  {Williams}}]{newhall83}
{Newhall}, X.~X., {Standish}, E.~M., \& {Williams}, J.~G. 1983, Astron.
  Astrophys., 125, 150

\bibitem[{{Oppenheimer} {et~al.}(2013){Oppenheimer}, {Baranec}, {Beichman},
  {Brenner}, {Burruss}, {Cady}, {Crepp}, {Dekany}, {Fergus}, {Hale},
  {Hillenbrand}, {Hinkley}, {Hogg}, {King}, {Ligon}, {Lockhart}, {Nilsson},
  {Parry}, {Pueyo}, {Rice}, {Roberts}, {Roberts}, {Shao}, {Sivaramakrishnan},
  {Soummer}, {Truong}, {Vasisht}, {Veicht}, {Vescelus}, {Wallace}, {Zhai}, \&
  {Zimmerman}}]{oppenheimer13}
{Oppenheimer}, B.~R., {Baranec}, C., {Beichman}, C., {et~al.} 2013,
  Astrophys.~J., 768, 24

\bibitem[{Poisson \& Will(2014)}]{poisson14}
Poisson, E., \& Will, C.~M. 2014, Gravity: Newtonian, Post-Newtonian,
  Relativistic ({Cambridge, pp.~780}: Cambridge University Press)

\bibitem[{{Quinn} {et~al.}(1991){Quinn}, {Tremaine}, \& {Duncan}}]{quinn91}
{Quinn}, T.~R., {Tremaine}, S., \& {Duncan}, M. 1991, Astron.~J., 101, 2287

\bibitem[{{Rauch} \& {Hamilton}(2002)}]{rauch02}
{Rauch}, K.~P., \& {Hamilton}, D.~P. 2002, in Bull. Am. Astron. Soc., Vol.~34,
  AAS/Division of Dynamical Astronomy Meeting \#33, 938

\bibitem[{{Rauch} \& {Holman}(1999)}]{rauch99}
{Rauch}, K.~P., \& {Holman}, M. 1999, Astron.~J., 117, 1087

\bibitem[{{Richter}(2001)}]{richter01}
{Richter}, P.~H. 2001, in Reviews in Modern Astronomy: Dynamic Stability and
  Instabilities in the Universe, ed. R.~E. {Schielicke}, Vol.~14, 53--92

\bibitem[{{Rushby} {et~al.}(2013){Rushby}, {Claire}, {Osborn}, \&
  {Watson}}]{rushby13}
{Rushby}, A.~J., {Claire}, M.~W., {Osborn}, H., \& {Watson}, A.~J. 2013,
  Astrobiology, 13, 833

\bibitem[{{Saha} \& {Tremaine}(1992)}]{saha92}
{Saha}, P., \& {Tremaine}, S. 1992, Astron.~J., 104, 1633

\bibitem[{{Schr{\"o}der} \& {Smith}(2008)}]{schroeder08}
{Schr{\"o}der}, K.-P., \& {Smith}, R.~C. 2008, Mon. Not. R. Astron. Soc., 386,
  155

\bibitem[{{Soffel}(1989)}]{soffel89}
{Soffel}, M.~H. 1989, Relativity in Astrometry, Celestial Mechanics and Geodesy
  (Springer-Verlag Heidelberg, pp.~208)

\bibitem[{{Sussman} \& {Wisdom}(1992)}]{sussman92}
{Sussman}, G.~J., \& {Wisdom}, J. 1992, Science, 257, 56

\bibitem[{{Varadi} {et~al.}(2003){Varadi}, {Runnegar}, \& {Ghil}}]{varadi03}
{Varadi}, F., {Runnegar}, B., \& {Ghil}, M. 2003, Astrophys.~J., 592, 620

\bibitem[{{Wisdom} \& {Holman}(1991)}]{wisdom91}
{Wisdom}, J., \& {Holman}, M. 1991, Astron.~J., 102, 1528

\bibitem[{Zeebe(2015)}]{zeebe15apj}
Zeebe, R.~E. 2015, Astrophys.~J., 798, 8

\end{thebibliography}
%


\begin{table}[hhhhhh]
\begin{center}
\caption{Initial conditions of the eight planets and Pluto$^a$ 
for 5-Gyr runs from DE431. Heliocentric positions \v{x} (AU) and 
velocities \v{v} (AU~day\pmo). \label{TabDE431}}
\begin{tabular}{lccc}
\tableline\tableline
 & $x$ & $y$ & $z$ \\
\tableline
 & & Mercury & \\
\v{x} &
 $-$1.40712354144735680E-01 & $-$4.43906230277241465E-01 & $-$2.33474338281349329E-02 \\
\v{v} &
  +2.11691765462179472E-02 & $-$7.09701275933066148E-03 & $-$2.52278032052283448E-03 \\
 & & Venus & \\
\v{x} &
 $-$7.18629835259113170E-01 & $-$2.25188858612526514E-02 & +4.11716131772919824E-02  \\
\v{v} &
  +5.13955712094533914E-04 & $-$2.03061283748202266E-02 & $-$3.07198741951420558E-04 \\
 & & Earth + Moon & \\
\v{x} &
 $-$1.68563248623229384E-01 &  +9.68761420122898564E-01 & $-$1.15183154209270563E-06 \\
\v{v} &
 $-$1.72299715055074729E-02 & $-$3.01349780674632205E-03 & +2.41254068070491868E-08 \\
 & & Mars & \\
\v{x} &
  +1.39036162161402177E+00 & $-$2.09984400533893799E-02 & $-$3.46177919349353047E-02 \\
\v{v} &
  +7.47813544105227729E-04 & +1.51863004086334515E-02 &  +2.99756038504512547E-04 \\
 & & Jupiter & \\
\v{x} &
  +4.00345668418424960E+00 &  +2.93535844833712467E+00 & $-$1.01823217020834328E-01 \\
\v{v} &
 $-$4.56348056882991196E-03 & +6.44675255807273997E-03 &  +7.54565159392195741E-05 \\
 & & Saturn & \\
\v{x} &
  +6.40855153734800886E+00 &  +6.56804703677062207E+00 & $-$3.69127809402511886E-01 \\
\v{v} &
 $-$4.29112154163879215E-03 &  +3.89157880254167561E-03 &  +1.02876894772680478E-04 \\
 & & Uranus & \\
\v{x} &
  +1.44305195077618524E+01 & $-$1.37356563056406209E+01 & $-$2.38128487167790809E-01 \\
\v{v} &
  +2.67837949019966498E-03 &  +2.67244291355153403E-03 & $-$2.47764637737944378E-05 \\
 & & Neptune & \\
\v{x} &
  +1.68107582839480649E+01 & $-$2.49926499733276124E+01 &  +1.27271208982211476E-01 \\
\v{v} &
  +2.57936917068014599E-03 &  +1.77676956230748452E-03 & $-$9.59089132565213410E-05 \\
 & & Pluto & \\
\v{x} &
 $-$9.87686582399026491E+00 & $-$2.79580297772433077E+01 &  +5.85080284687055574E+00 \\
\v{v} &
  +3.02870206449818878E-03 & $-$1.53793257901232473E-03 & $-$7.12171623386267461E-04 \\
\tableline
\end{tabular}
\tablenotetext{a}{Masses (Mercury to Pluto in solar masses):
1.66013679527193035E-07,
2.44783833966454472E-06,
3.04043264626852573E-06,
3.22715144505387430E-07,
9.54791938424322164E-04,
2.85885980666102893E-04,
4.36625166899970042E-05,
5.15138902053549668E-05,
7.40740740740740710E-09.
}
\end{center}
\end{table}

\end{document}